\documentclass[manyauthors,nocleardouble,COMPASS]{cernphprep}

\usepackage{bm}
\usepackage{color}

\usepackage{amssymb}
\usepackage{amsmath}
\usepackage{amsbsy}
\usepackage{times}
\usepackage{epsfig}
\usepackage{colordvi}
\usepackage{graphicx}
\usepackage{wrapfig,rotating}
\usepackage[numbers, square, comma, sort&compress]{natbib}
\usepackage{hyperref} 
\usepackage{multicol}
\usepackage{booktabs}
\usepackage{multirow}
\usepackage{lineno}
\RequirePackage[T1]{fontenc}
\makeatletter

\DeclareMathOperator{\Tr}{Tr}
\DeclareSymbolFont{letters}     {OML}{cmm}{m}{it}
\DeclareSymbolFont{symbols}     {OMS}{cmsy}{m}{n}
\DeclareSymbolFont{largesymbols}{OMX}{cmex}{m}{n}
\begin{document}

\begin{titlepage}

\PHnumber{2014--096}
\PHdate{May 16, 2014}

\title{Spin alignment and violation of the OZI rule in exclusive $\omega$ and
  $\phi$ production in pp collisions} 

\Collaboration{The COMPASS Collaboration}
\ShortAuthor{The COMPASS Collaboration}

\begin{abstract}
Exclusive production of the isoscalar vector mesons $\omega$ and $\phi$ is
measured with a 190\,GeV$/c$ proton beam impinging on a liquid hydrogen
target. Cross section ratios are determined in three intervals of the Feynman
variable $x_{F}$ of the fast proton. A significant violation of the OZI rule is
found, confirming earlier findings. Its kinematic dependence on $x_{F}$ and on
the invariant mass $M_{p\mathrm{V}}$ of the system formed by fast proton
$p_\mathrm{fast}$ and vector meson $V$ is discussed in terms of diffractive
production of $p_\mathrm{fast}V$ resonances in competition with central
production. The measurement of the spin density matrix element $\rho_{00}$ of
the vector mesons in different selected reference frames provides another handle
to distinguish the contributions of these two major reaction types. Again,
dependences of the alignment on $x_{F}$ and on $M_{p\mathrm{V}}$ are found. Most
of the observations can be traced back to the existence of several excited
baryon states contributing to $\omega$ production which are absent in the case
of the $\phi$ meson. Removing the low-mass $M_{p\mathrm{V}}$ resonant region,
the OZI rule is found to be violated by a factor of eight, independently of
$x_\mathrm{F}$.\\


\end{abstract}
\vfill
\Submitted{(to be submitted to Nucl. Phys. B)}

\end{titlepage}

{\pagestyle{empty}
%%%%%%%%%%%%%%%%%%%%%%%%%%%%%%%%%%%%%%%%%%%%%%%%%%%%%%%%%%%%%%%%%%%%%%%%%%%%%%
%
% 2014_auththorlist.tex  
%
%%%%%%%%%%%%%%%%%%%%%%%%%%%%%%%%%%%%%%%%%%%%%%%%%%%%%%%%%%%%%%%%%%%%%%%%%%%%%%

\section*{The COMPASS Collaboration}
\label{app:collab}
\renewcommand\labelenumi{\textsuperscript{\theenumi}~}
\renewcommand\theenumi{\arabic{enumi}}
\begin{flushleft}
C.~Adolph\Irefn{erlangen},
R.~Akhunzyanov\Irefn{dubna}, %phd
M.G.~Alexeev\Irefn{turin_u},
G.D.~Alexeev\Irefn{dubna}, %1
A.~Amoroso\Irefnn{turin_u}{turin_i},
V.~Andrieux\Irefn{saclay},
V.~Anosov\Irefn{dubna}, %2
A.~Austregesilo\Irefnn{cern}{munichtu},
B.~Bade{\l}ek\Irefn{warsawu},
F.~Balestra\Irefnn{turin_u}{turin_i}
% ...
\end{flushleft}

%%%%%%%%%%%%%%%%%%%%%%%%%%%%%%%%%%%%%%%%%%%%%%%%%%%%%%%%%%%%%%%%%%%%%%%%%%%%%%%%%%%%%%%%%%%%%%%%%%%%%%%%%%%%%%%%%%%%%%%
%
% institutes
%
%%%%%%%%%%%%%%%%%%%%%%%%%%%%%%%%%%%%%%%%%%%%%%%%%%%%%%%%%%%%%%%%%%%%%%%%%%%%%%%%%%%%%%%%%%%%%%%%%%%%%%%%%%%%%%%%%%%%%%%

\begin{Authlist}
\item \Idef{bielefeld}{Universit\"at Bielefeld, Fakult\"at f\"ur Physik, 33501 Bielefeld, Germany\Arefs{f}}
\item \Idef{bochum}{Universit\"at Bochum, Institut f\"ur Experimentalphysik, 44780 Bochum, Germany\Arefs{f}\Arefs{ll}}
\item \Idef{bonniskp}{Universit\"at Bonn, Helmholtz-Institut f\"ur  Strahlen- und Kernphysik, 53115 Bonn, Germany\Arefs{f}}
\item \Idef{bonnpi}{Universit\"at Bonn, Physikalisches Institut, 53115 Bonn, Germany\Arefs{f}}
\item \Idef{brno}{Institute of Scientific Instruments, AS CR, 61264 Brno, Czech Republic\Arefs{g}}
\item \Idef{calcutta}{Matrivani Institute of Experimental Research \& Education, Calcutta-700 030, India\Arefs{h}}
\item \Idef{dubna}{Joint Institute for Nuclear Research, 141980 Dubna, Moscow region, Russia\Arefs{i}}
\item \Idef{erlangen}{Universit\"at Erlangen--N\"urnberg, Physikalisches Institut, 91054 Erlangen, Germany\Arefs{f}}
\item \Idef{freiburg}{Universit\"at Freiburg, Physikalisches Institut, 79104 Freiburg, Germany\Arefs{f}\Arefs{ll}}
\item \Idef{cern}{CERN, 1211 Geneva 23, Switzerland}
\item \Idef{liberec}{Technical University in Liberec, 46117 Liberec, Czech Republic\Arefs{g}}
\item \Idef{lisbon}{LIP, 1000-149 Lisbon, Portugal\Arefs{j}}
\item \Idef{mainz}{Universit\"at Mainz, Institut f\"ur Kernphysik, 55099 Mainz, Germany\Arefs{f}}
\item \Idef{miyazaki}{University of Miyazaki, Miyazaki 889-2192, Japan\Arefs{k}}
\item \Idef{moscowlpi}{Lebedev Physical Institute, 119991 Moscow, Russia}
\item \Idef{munichlmu}{Ludwig-Maximilians-Universit\"at M\"unchen, Department f\"ur Physik, 80799 Munich, Germany\Arefs{f}\Arefs{l}}
\item \Idef{munichtu}{Technische Universit\"at M\"unchen, Physik Department, 85748 Garching, Germany\Arefs{f}\Arefs{l}}
\item \Idef{nagoya}{Nagoya University, 464 Nagoya, Japan\Arefs{k}}
\item \Idef{praguecu}{Charles University in Prague, Faculty of Mathematics and Physics, 18000 Prague, Czech Republic\Arefs{g}}
\item \Idef{praguectu}{Czech Technical University in Prague, 16636 Prague, Czech Republic\Arefs{g}}
\item \Idef{protvino}{State Scientific Center Institute for High Energy Physics of National Research Center `Kurchatov Institute', 142281 Protvino, Russia}
\item \Idef{saclay}{CEA IRFU/SPhN Saclay, 91191 Gif-sur-Yvette, France\Arefs{ll}}
\item \Idef{telaviv}{Tel Aviv University, School of Physics and Astronomy, 69978 Tel Aviv, Israel\Arefs{m}}
\item \Idef{triest_u}{University of Trieste, Department of Physics, 34127 Trieste, Italy}
\item \Idef{triest_i}{Trieste Section of INFN, 34127 Trieste, Italy}
\item \Idef{triest_ictp}{Abdus Salam ICTP, 34151 Trieste, Italy}
\item \Idef{turin_u}{University of Turin, Department of Physics, 10125 Turin, Italy}
\item \Idef{turin_p}{University of Eastern Piedmont, 15100 Alessandria, Italy}
\item \Idef{turin_i}{Torino Section of INFN, 10125 Turin, Italy}
\item \Idef{warsaw}{National Centre for Nuclear Research, 00-681 Warsaw, Poland\Arefs{n} }
\item \Idef{warsawu}{University of Warsaw, Faculty of Physics, 00-681 Warsaw, Poland\Arefs{n} }
\item \Idef{warsawtu}{Warsaw University of Technology, Institute of Radioelectronics, 00-665 Warsaw, Poland\Arefs{n} }
\item \Idef{yamagata}{Yamagata University, Yamagata, 992-8510 Japan\Arefs{k} }
\end{Authlist}
%%%%%%%%%%%%%%%%%%%%%%%%%%%%%%%%%%%%%%%%%%%%%%%%%%%%%%%%%%%%%%%%%%%%%%%%%%%%%%%%%%%%%%%%%%%%%%%%%%%%%%%%%%%%%%%%%%%%%%%
%
% Notes
%
%%%%%%%%%%%%%%%%%%%%%%%%%%%%%%%%%%%%%%%%%%%%%%%%%%%%%%%%%%%%%%%%%%%%%%%%%%%%%%%%%%%%%%%%%%%%%%%%%%%%%%%%%%%%%%%%%%%%%%%
\vspace*{-\baselineskip}\renewcommand\theenumi{\alph{enumi}}
\begin{Authlist}
\item \Adef{a}{Also at Instituto Superior T\'ecnico, Universidade de Lisboa, Lisbon, Portugal}
\item \Adef{aa}{Also at Department of Physics, Pusan National University, Busan 609-735, Republic of Korea and at Physics Department, Brookhaven National Laboratory, Upton, NY 11973, U.S.A. }
\item \Adef{bb}{Supported by the DFG Research Training Group Programme 1102  ``Physics at Hadron Accelerators''}
\item \Adef{b}{Also at Chubu University, Kasugai, Aichi, 487-8501 Japan\Arefs{k}}
\item \Adef{c}{Also at KEK, 1-1 Oho, Tsukuba, Ibaraki, 305-0801 Japan}
\item \Adef{c1c}{Present address: Universit\"at Bonn, Helmholtz-Institut f\"ur Strahlen- und Kernphysik, 53115 Bonn, Germany}
\item \Adef{cc}{Also at Moscow Institute of Physics and Technology, Moscow Region, 141700, Russia}
\item \Adef{x}{present address: RWTH Aachen University, III. Physikalisches Institut, 52056 Aachen, Germany}
\item \Adef{f}{Supported by the German Bundesministerium f\"ur Bildung und Forschung}
\item \Adef{g}{Supported by Czech Republic MEYS Grants ME492 and LA242}
\item \Adef{h}{Supported by SAIL (CSR), Govt.\ of India}
\item \Adef{i}{Supported by CERN-RFBR Grants 08-02-91009 and 12-02-91500}
\item \Adef{j}{\raggedright Supported by the Portuguese FCT - Funda\c{c}\~{a}o para a Ci\^{e}ncia e Tecnologia, COMPETE and QREN, Grants CERN/FP/109323/2009, CERN/FP/116376/2010 and CERN/FP/123600/2011}
\item \Adef{k}{Supported by the MEXT and the JSPS under the Grants No.18002006, No.20540299 and No.18540281; Daiko Foundation and Yamada Foundation}
\item \Adef{l}{Supported by the DFG cluster of excellence `Origin and Structure of the Universe' (www.universe-cluster.de)}
\item \Adef{ll}{Supported by EU FP7 (HadronPhysics3, Grant Agreement number 283286)}
\item \Adef{m}{Supported by the Israel Science Foundation, founded by the Israel Academy of Sciences and Humanities}
\item \Adef{n}{Supported by the Polish NCN Grant DEC-2011/01/M/ST2/02350}
\item [{\makebox[2mm][l]{\textsuperscript{*}}}] Deceased
\end{Authlist}

\newpage

\section{Introduction}
\label{sec:Introduction}
The Okubo-Zweig-Iizuka (OZI) rule\,\cite{OZI} was formulated in the early days
of the quark model, stating that all hadronic processes with disconnected quark
lines are suppressed.  It qualitatively explains phenomena like suppression of
$\phi$ meson decays into non-strange particles and suppression of exclusive
$\phi$ production in non-strange hadron collisions. Using the known deviation
from the ideal mixing angle of the vector mesons $\omega$ and $\phi$,
$\delta_\mathrm{V} = 3.7^{\mathrm{o}}$, the production cross section of $\phi$
with respect to that of $\omega$ should be suppressed according to $\sigma(AB
\rightarrow X\phi)/\sigma(AB \rightarrow X\omega) = \tan^2\delta_\mathrm{V} =
0.0042$, where $A$, $B$ and $X$ are non-strange hadrons \cite{lipkin}. At low
energies, where baryonic and mesonic degrees of freedom are most relevant, the
ratio can be expressed in terms of meson-meson or meson-nucleon couplings:
$g^2_{\phi\rho\pi}/g^2_{\omega\rho\pi} = g^2_{\phi NN}/g^2_{\omega NN} =
\tan^2\delta_\mathrm{V} = 0.0042$, where $N$ denotes the nucleon. This is valid
provided the coupling ratios $g_{\phi\rho\pi}/g_{\omega\rho\pi}$ and $g_{\phi
  NN}/g_{\omega NN}$ are equal as advocated in Ref.\,\cite{lipkin2}.

The OZI rule was tested in several experiments and is remarkably well fulfilled
in many reactions (for a review, see \textit{e.g.} Refs.\,\cite{sapozhnikov} and
\cite{sibirtsev}).  Apparent violations of the OZI rule -- observed in
$p\bar{p}$ annihilations at rest and in nucleon-nucleon collisions -- can be
interpreted either as a true violation due to gluonic intermediate states (see
\textit{e.g.} Ref.\,\cite{lindenbaum}) or as an evasion from the OZI rule
because of a hidden strangeness component in the nucleon \cite{ellis}.  Such a
strangeness component, possibly polarised, was suggested as an explanation of
the apparent OZI violations observed in
$p\,N\,\rightarrow\,N\,p\,V,~V\,=\,\omega,\phi$ by the SPHINX
collaboration\,\cite{sphinx1}.  Large OZI violations at low energies have also
led to speculations about crypto-exotic baryon resonances decaying to $N\phi$
\cite{sibirtsevcrypto}.

Although being phenomenological in its origin, the OZI rule has been connected
to QCD \cite{lipkin}. In a field theoretical approach to the OZI rule, a
perturbative treatment based on quark-gluon degrees of freedom requires the
scale of a specific process to be much larger than the QCD cut-off parameter
$\Lambda_{QCD}\approx200$\,MeV/$c$.  In charmonium production, where the scale
is governed by the charm quark current mass $m_{c}\approx1275$\,MeV/$c^2$, the
quark--antiquark pair is generated by gluon splitting, $g\rightarrow c\bar{c}$.
This is in contrast to the case of strangeness production, where the scale
corresponds to the strange quark current mass $m_{s}\approx95$\,MeV/$c^2$, which
is close to $\Lambda_{QCD}$. The validity of the quark-gluon picture can thus be
questioned, and the relevant degrees of freedom need to be determined. Gluon
splitting can only be used in an effective sense. This has also been discussed
in connection to hyperon production in $\bar{p}p \rightarrow
\bar{\Lambda}\Lambda$ production near threshold, where neither meson exchange
models nor quark-gluon models give a complete explanation of the experimental
data\,\cite{klempt}. However, probed at virtualities $Q^{2}$ or
$p_{\perp}^{2}\gg1$\,(GeV$/c)^2$, which are large compared to
$(2m_{s})^{2}c^2\approx\Lambda_{QCD}^2\approx0.04$\,(GeV$/c)^2$, the process can
be described in the quark-gluon picture and we expect strangeness suppression to
disappear, restoring flavour SU(3) symmetry.

In this work, we present an attempt to understand the effective scale governing
the (hidden) strangeness production in the exclusive process
$p\,p\,\rightarrow\,p\,\phi\,p$ by studying the degree of OZI violation.  The
difficulty lies in the separation of different reaction mechanisms as a function
of transferred energy and angular momentum. The latter is reflected in the
anisotropy of the decay angular distributions which can be expressed
\textit{via} the spin density matrix\,\cite{gj}.  In the analysis of data from
an unpolarised beam impinging on an unpolarised target, symmetries leave one
independent element of the spin density matrix, $\rho_{00}$, which is a measure
for spin alignment (tensor polarisation). It can be extracted from distributions
of the angle between the decay plane (3-body decay) or decay axis (2-body decay)
of the vector meson and a well-chosen reference axis \cite{schilling}.

The MOMO collaboration measured $\rho_{00}$ of the $\phi$ meson in
$p\,d\,\rightarrow\,^{3}$He$\,\phi$ near the kinematic threshold and the result
was consistent with a complete alignment of the $\phi$ meson with respect to the
incoming beam\,\cite{momo}. This is in sharp contrast to the case of the
$\omega$ meson, which is produced unaligned at the same excess energy and in the
same initial state, as found by the WASA collaboration\,\cite{wasa}. The
alignment of the $\omega$ meson in $pp$ collisions was measured close to
threshold by the COSY-TOF collaboration\,\cite{anke} and in $pN$ collisions at a
beam momentum of 70\,GeV/$c$ by SPHINX\,\cite{sphinx2}, whereas the $\phi$
alignment was measured at high energies by ACCMOR\,\cite{accmor} and by STAR at
RHIC\,\cite{rhic}. Prior to our measurement, the only simultaneous measurement
of $\phi$ and $\omega$ alignment using the same experimental set-up was
performed by the SAPHIR collaboration\,\cite{saphir1,saphir2} in
photoproduction.

At COMPASS, the exclusive reaction
$p_\mathrm{beam}\,p_\mathrm{target}\,\rightarrow\,p_\mathrm{fast}\,V\,
p_\mathrm{recoil}$ is measured at a beam momentum of 190\,GeV/$c$. For
simplicity, this will from now on be denoted $p\,p\,\rightarrow\,p\,V\,
p$. Apart from this notation and unless otherwise stated explicitly, the symbol
$p$ without subscript and the Feynman variable $x_\mathrm{F}=
p_L/p_{L\mathrm{max}}$, $p_L$ denoting the longitudinal momentum, will refer to
the fast proton. The reduced 4-momentum transfer squared $t'$ from the beam to
the recoil proton is defined as $t'=|t|-|t|_\mathrm{min}$, where
$t=\left(p_{p\mathrm{beam}}-(p_{p\mathrm{fast}}+p_{V})\right)^2$ and
$|t|_\mathrm{min}$ the minimum value of $|t|$.

For exclusive vector meson production, there are contributions from mainly two
classes of processes: resonant and non-resonant production. First, resonant
production means diffractive dissociation of the fast proton, where a Pomeron is
emitted in the $t$-channel from the target and excites the beam particle (see
Fig.\,\ref{fig:prodmech}, left panel). The target particle receives a small
recoil but stays intact. The vector meson is then produced \textit{via} a baryon
resonance. On the other side, there is the non-resonant process including the
case when a vector meson is radiated from the proton in the initial or final
state. This is possible due to a finite coupling of the vector meson to the
meson cloud of the nucleon \cite{titov}. These non-resonant processes are
summarised in the middle panel of Fig.\,\ref{fig:prodmech}, where the blob in
the upper vertex represents point-like and non-point-like
interactions. Non-resonant vector meson production also includes central
production where a Reggeon or Pomeron from the target and a Reggeon or Pomeron
from the beam particle fuse in a central vertex (see Fig.\,\ref{fig:prodmech},
right panel). The production of $\omega$ and $\phi$ in Pomeron-Pomeron
collisions does not conserve $G$-parity and is thus forbidden. Central
Production is characterised by large rapidity gaps between all three final state
particles. This is equivalent to large gaps between the $x_\mathrm{F}$
distributions of the outgoing particles. For the $p\,p\,\rightarrow\,p\,V\, p$
process this results in large $x_\mathrm{F}$ of the fast proton. Another special
case of non-resonant production is the shake-out (see \textit{e.g.}
Ref. \cite{ellis}) of a $q\overline{q}$ pair from the sea of one nucleon which
becomes on-shell when interacting with a Pomeron from the other nucleon. In the
case of shake-out, a rapidity gap is expected between the recoil particle and
the other two particles, but not necessarily between the fast proton and the
vector meson. Central production and shake-out can in this sense be considered
as similar processes in two different regions of phase space.

The dynamics of the vector meson is determined by the incoming particles of the
production vertex. In the case of Pomeron--Reggeon fusion and shake-out, the
dynamics of the vector meson depends on the exchange object(s) while in resonant
diffractive production, it depends on the intermediate resonance.

\begin{figure}[h!]
  \centering
  \includegraphics[width=\textwidth]{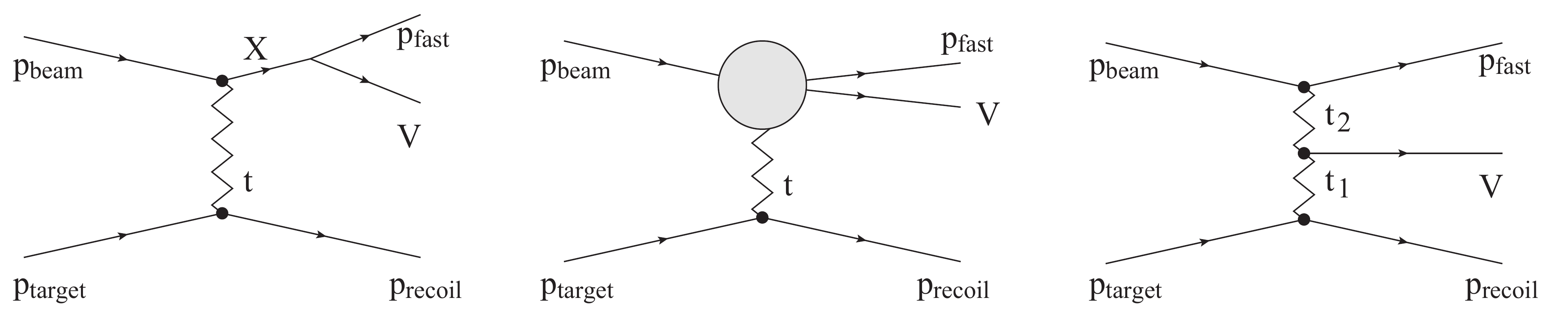}
  \caption[Production mechanisms]{Mechanisms for exclusive vector meson
    production at high energies. Left: Resonant single diffractive dissociation
    of the beam proton to a resonance X with subsequent decay. Middle:
    Non-resonant single diffractive excitation of the beam proton. The blob in
    the upper vertex denotes both point-like and non-point-like
    interactions. Right: Central production.} 
  \label{fig:prodmech}
\end{figure}

In this work, the cross section ratio 
\begin{equation}
R_{\phi/\omega} = \frac{d\sigma(p\,p\,\rightarrow\,p\,\phi\,p)/dx_\mathrm{F}}{d\sigma(p\,p\,\rightarrow\,p\,\omega\,p)/dx_\mathrm{F}}
\label{eq:Rphiomega}
\end{equation}
 is presented as a function of $x_\mathrm{F}$ using different constraints on the
 invariant mass of proton and vector meson, $M_{p\mathrm{V}}$. The data are in
 the kinematic domain $0<p_{\perp}^{2}<1$\,(GeV$/c)^2$. We also study the spin
 alignment of $\omega$ and $\phi$ and its dependence on $x_\mathrm{F}$ and
 $M_{p\mathrm{V}}$ in different reference frames. 

\section{Experimental set-up}
The COMPASS experiment uses a fixed-target experiment situated at the M2 beam
line of the CERN SPS. A detailed description can be found in
Ref.\,\cite{compassnim}. For the present measurement, a beam of 190\,GeV/$c$
positively charged hadrons with a nominal intensity of $5 \cdot 10^6$\,s$^{-1}$
and a spill length of 10\,s every 45\,s was used. The positive beam is composed
of 74.6\% protons, 24.0\% pions and 1.4\% kaons. Each beam particle is
identified using two differential Cherenkov detectors (CEDAR) and its trajectory
is measured with a silicon microstrip telescope in front of the target.

The liquid hydrogen target with a length of 400\,mm and a diameter of 35\,mm is
surrounded by two cylindrical layers of scintillators (RPD) for time-of-flight
and ${\rm d}E/{\rm d}x$ measurements of the slow target-recoil protons.  The
material of the target, the vacuum pipe and the inner layer of the RPD imply a
minimum momentum transfer squared of $|t|=0.07$\,(GeV$/c)^2$ for detection of
recoil protons.

The other final state particles are detected in a two-stage open forward
spectrometer with large acceptance in momentum and angle. The small acceptance
gap between the RPD and the forward spectrometer is covered by a
lead-scintillator sandwich detector used as veto.  The first and second
spectrometer stage consists of a dipole magnet surrounded by tracking detectors
followed by electromagnetic (ECAL1 and ECAL2) and hadron calorimeters. The first
stage also contains a ring-imaging Cherenkov counter (RICH) for pion/kaon
separation up to 50\,GeV/$c$. Using C$_4$F$_{10}$ as radiator gas, thresholds of
2.5\,GeV/$c$ and 9\,GeV/$c$ are obtained for pions and kaons, respectively.

The trigger system selects interactions in the target material by requiring a
recoil proton in addition to an incoming beam particle. These requirements avoid
any influence of the trigger onto the selection of particles in the forward
spectrometer.

\section{Analysis}
\subsection{Event selection}
\label{sec:eventselection}
The results presented in this paper are obtained by selecting $\omega$ and
$\phi$ mesons from the reactions
$pp\,\rightarrow\,p\,\omega\,p,\,\omega\,\rightarrow\,\pi^+\pi^-\pi^0$ and
$pp\,\rightarrow\,p\,\phi\,p,\,\phi\,\rightarrow\,K^+K^-$, respectively. The
data were taken in 2008 and 2009 and correspond to an integrated luminosity of
about 0.9\,pb$^{-1}$.

Exactly one well-defined interaction vertex is required to be reconstructed
within the target volume, for which the total charge of the three outgoing
charged tracks is +1. The incoming beam particle must be identified as a proton
in the CEDAR detectors. Furthermore, only events with exactly one proton
detected in the RPD are selected.

For the selection of a $\pi^0$ in the $\omega\,\rightarrow\,\pi^+\pi^-\pi^0$
channel, at least two photon candidates are required, defined as neutral
clusters in ECAL1 or ECAL2 with no associated reconstructed tracks. Energy
thresholds of $1$\,GeV and $2$\,GeV are applied to ECAL1 and ECAL2,
respectively. Furthermore, we require a photon pair in each event with invariant
mass within a window around the $\pi^0$ PDG value, which corresponds to $\pm
2\sigma_\mathrm{ECAL}$, where $\sigma_\mathrm{ECAL}$ is the mass resolution of a
photon pair in the calorimeter. The momentum of the $\pi^0$ is then recalculated
using a fit constrained to the PDG $\pi^0$ mass value to improve the
resolution. The $\pi^+$ must be identified in the RICH detector. The separation
of kaons and pions is done \textit{via} a log-likelihood method. The likelihood
for a pion hypothesis for the measured particle is required to be larger than
the likelihood for all other possible particle assignments. Furthermore, RICH
efficiencies are used to correct the particle yields. The sum of energies of the
final state particles detected in the spectrometer must be within a window of
$\pm$\,5\,GeV around the beam energy of 191\,GeV, referred to in the following
as exclusivity condition. The azimuthal angle of the forward going system
($\pi^+\pi^-\pi^0$ and the fast proton) and the azimuthal angle of the recoil
proton must differ by 180$^\circ$ within a window of $\pm\,16^\circ$
(coplanarity), which corresponds to twice the angular resolution of the RPD.

For the selection of $\phi$ mesons, the $K^+$ must be identified in the RICH
detector. Kaons are identified within a smaller momentum range than pions by the
RICH which imposes a momentum cut of about $10-50$\,GeV$/c$ on kaons and
influences the acceptance (see Sec.\,\ref{sec:acceptance}). In order to accept a
measured particle as a kaon, the likelihood for the kaon hypothesis must be 1.3
times larger than the likelihood obtained by any other possible particle
assignment including background. Again, RICH efficiencies are used to correct
the particle yields. Exclusivity and coplanarity are required as in the case of
$\pi^+\pi^-\pi^0$.

The reduced four-momentum transfer squared $t'$ is limited to values larger than
0.1\,$(\mathrm{GeV}/c)^2$ due to the RPD acceptance. The invariant mass of the
system $pV$, denoted as $M_{p\mathrm{V}}$, is constrained to 1.8\,GeV$/c^2 <
M_{p\omega}< 4.0$\,GeV$/c^2$ and 2.1\,GeV$/c^2 < M_{p\phi} < 4.5$\,GeV$/c^2$.

\subsection{Acceptance}
\label{sec:acceptance}
The spectrometer acceptance is accounted for by using a Monte Carlo (MC) based
multi-dimensional correction. The Monte Carlo event generator assumes the
two-step process $pp \rightarrow p_\mathrm{recoil}X,~X \rightarrow pV$, where
the intermediate resonance $X$ decays to the fast proton $p$ and the vector
meson $V$ according to phase space and where the $t'$ dependence of $\exp(-6.5
t')$ and the minimum $t'=0.07\,(\mathrm{GeV}/c)^2$ are taken from real data. The
Monte Carlo events are generated in narrow bins in $M_X$, \textit{i.e.} the mass
of $X$, and the total generated $M_X$ range covers the COMPASS spectrometer
acceptance. A beam parameterisation obtained from real data is used as input to
the generator in order to achieve realistic beam conditions, including
horizontal and vertical divergence of the beam for any given position of the
interaction vertex.

The propagation of the generated particles and their decay products through the
COMPASS spectrometer is simulated by the software package COMGEANT based on
GEANT3\,\cite{geant3}. The efficiency and purity of the RICH detector are
parameterised using real data, for details see Ref.\,\cite{promme}.  In order to
achieve a model independent correction, we use a three-dimensional acceptance
matrix in $t'$, $M_{p\mathrm{V}}$ and $x_\mathrm{F}$ of the fast proton.  Each
$K^+K^-$ or $\pi^+\pi^-\pi^0$ event from the collected data set is weighted by
the corresponding entry in the three-dimensional cell ($t'$, $M_{p\mathrm{V}}$
and $x_\mathrm{F}$) of the acceptance matrix. In a different approach, the
results are re-calculated using a different acceptance matrix where
$x_\mathrm{F}$ is replaced by $\cos{\theta}$, with $\theta$ being the helicity
angle of the $pV$ system as defined in Sec.\,\ref{sec:spinhel}. The results
differ by less than 1\%.  The statistical uncertainty of each value of the
acceptance matrix stems from a binomial probability density function as
described in Ref.\,\cite{TErnst}. It is typically 3--5 times smaller than the
statistical error from the real data and hence neglected.

The upper panels of Fig.\,\ref{fig:acc_xf} depict the $x_\mathrm{F}$ projection
of the acceptance matrix for both final states. While the acceptance remains
sizeable for $\pi^+\pi^-\pi^0$ down to $x_\mathrm{F} = 0.2$, it changes more
rapidly for $K^+K^-$ due to the RICH detector. The analysis is therefore
restricted to $0.6 < x_\mathrm{F} < 0.9$ in both channels in order to compare
$\phi$ and $\omega$ production within the same kinematic range. The impact of
the acceptance correction on the uncorrected $x_\mathrm{F}$ distributions for
vector meson, recoil and fast proton (shown in the middle panels of
Fig.\,\ref{fig:acc_xf}) is seen in the corresponding acceptance-corrected
distributions (shown in the lower panels of Fig.\,\ref{fig:acc_xf}). Note, that
the latter only contain events for $0.6 < x_\mathrm{F} < 0.9$, as described
above. Note the clear peaks for high $x_\mathrm{F}(p_\mathrm{fast})$ and small
$x_\mathrm{F}(\phi)$ distributions, indicating a contribution from central
production.
\begin{figure}[h!]
  \centering
  \includegraphics[width=.45\textwidth]{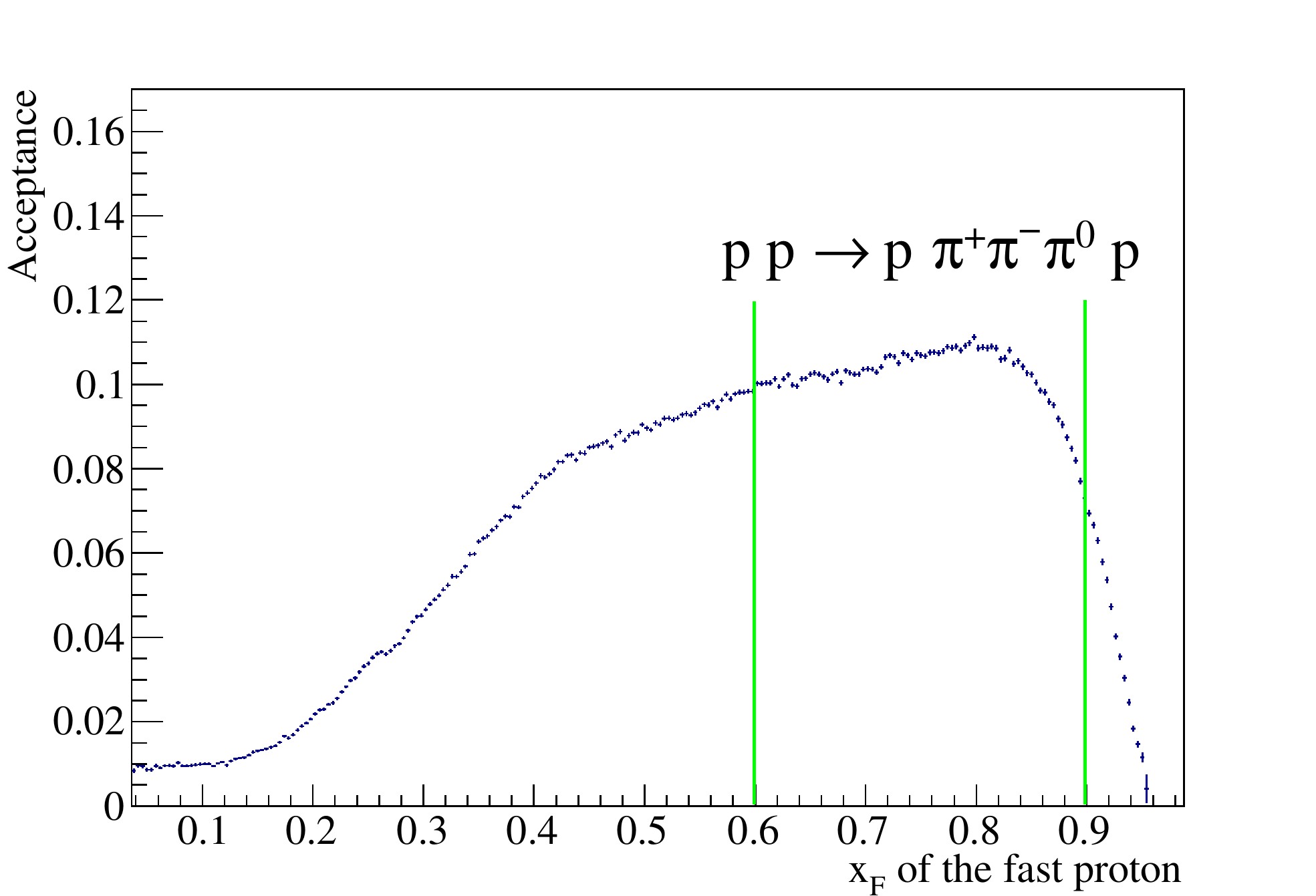}
  \includegraphics[width=.45\textwidth]{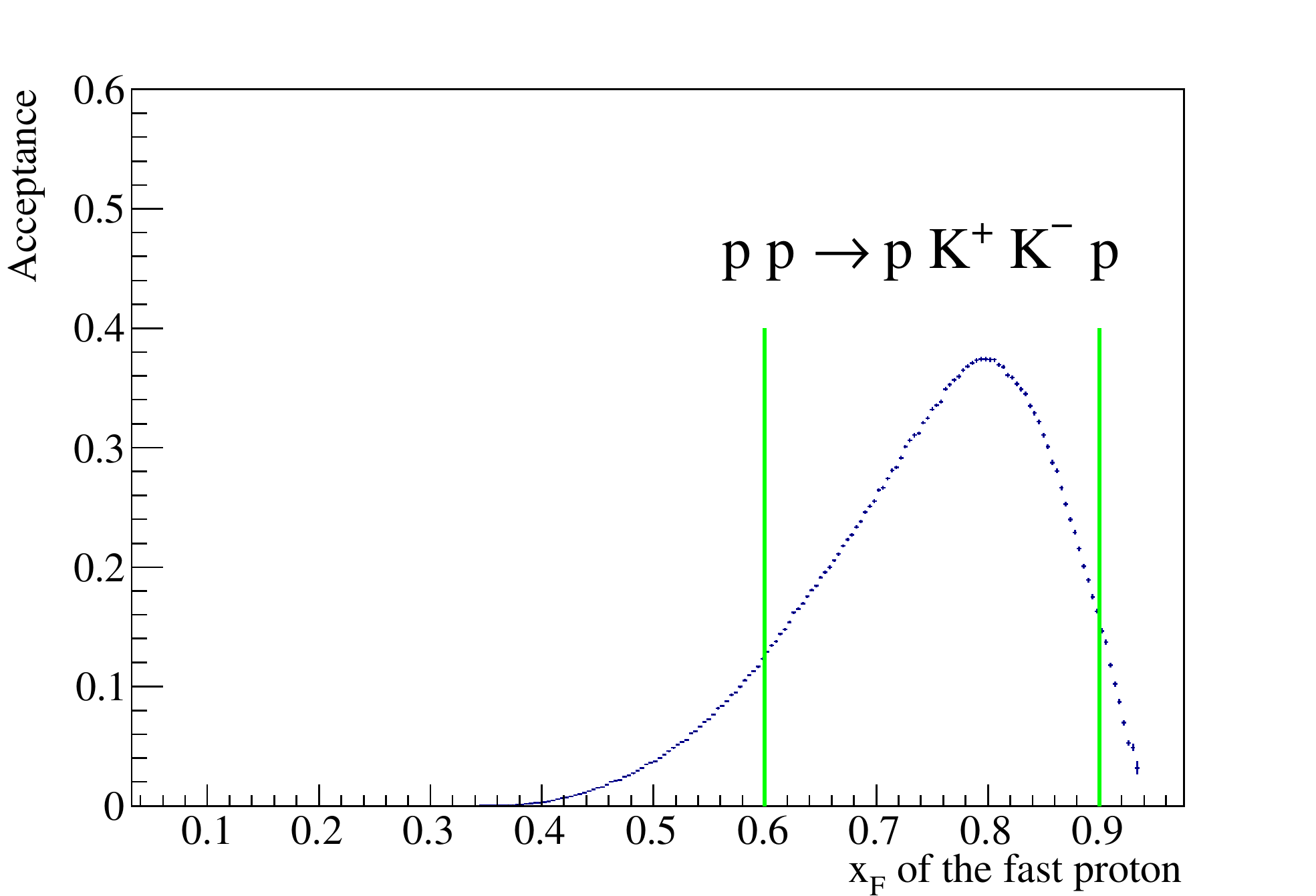}
  \includegraphics[width=.45\textwidth]{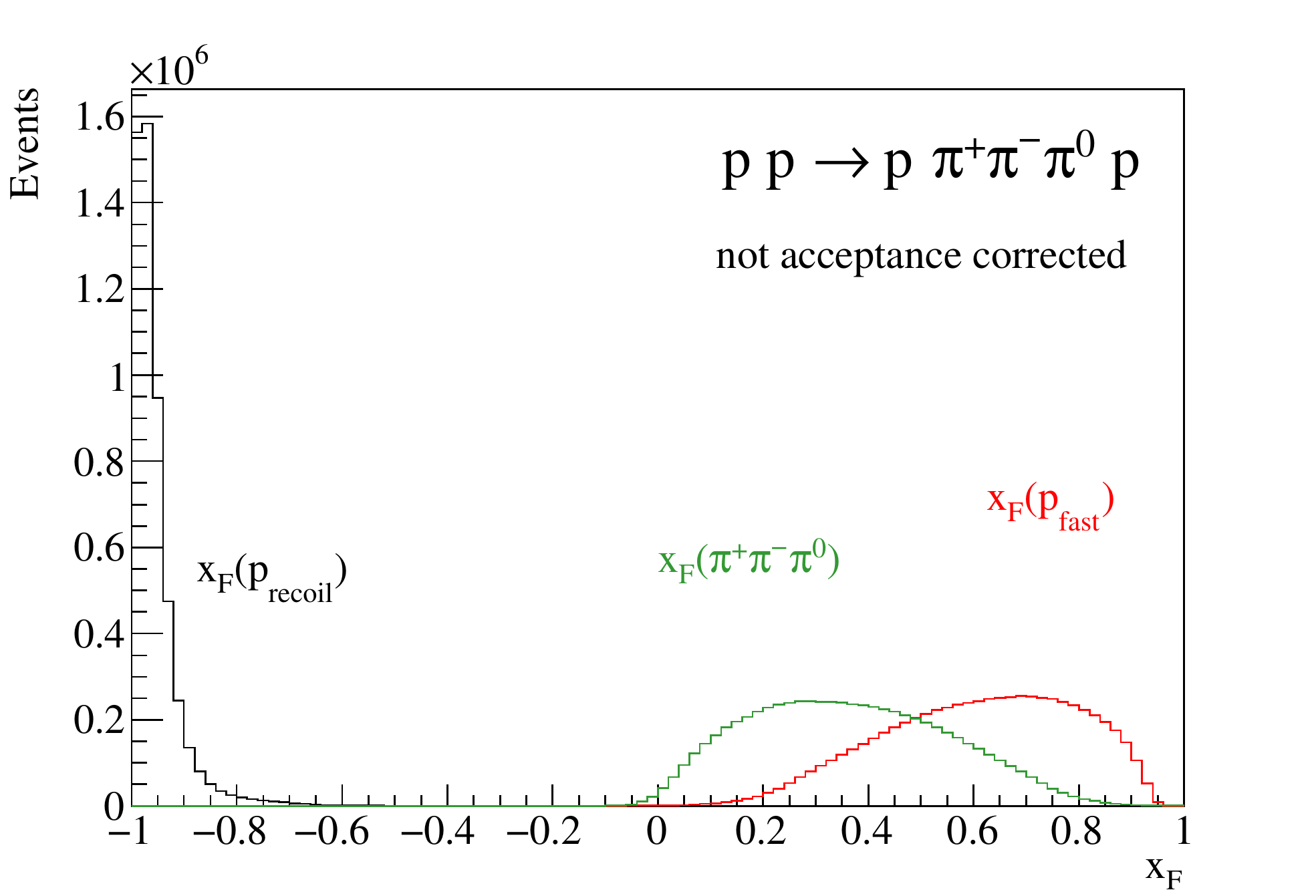}
  \includegraphics[width=.45\textwidth]{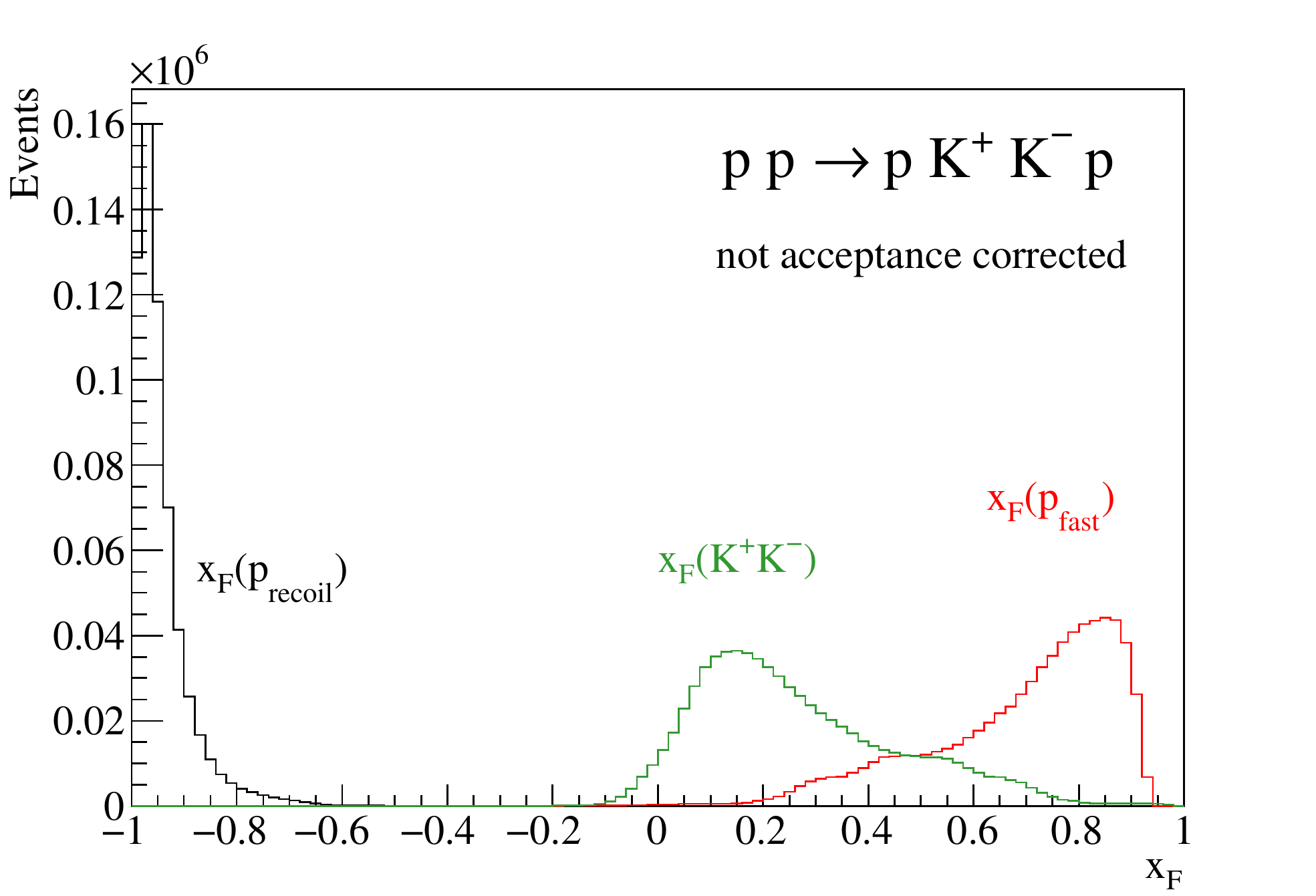}\\
  \includegraphics[width=.45\textwidth]{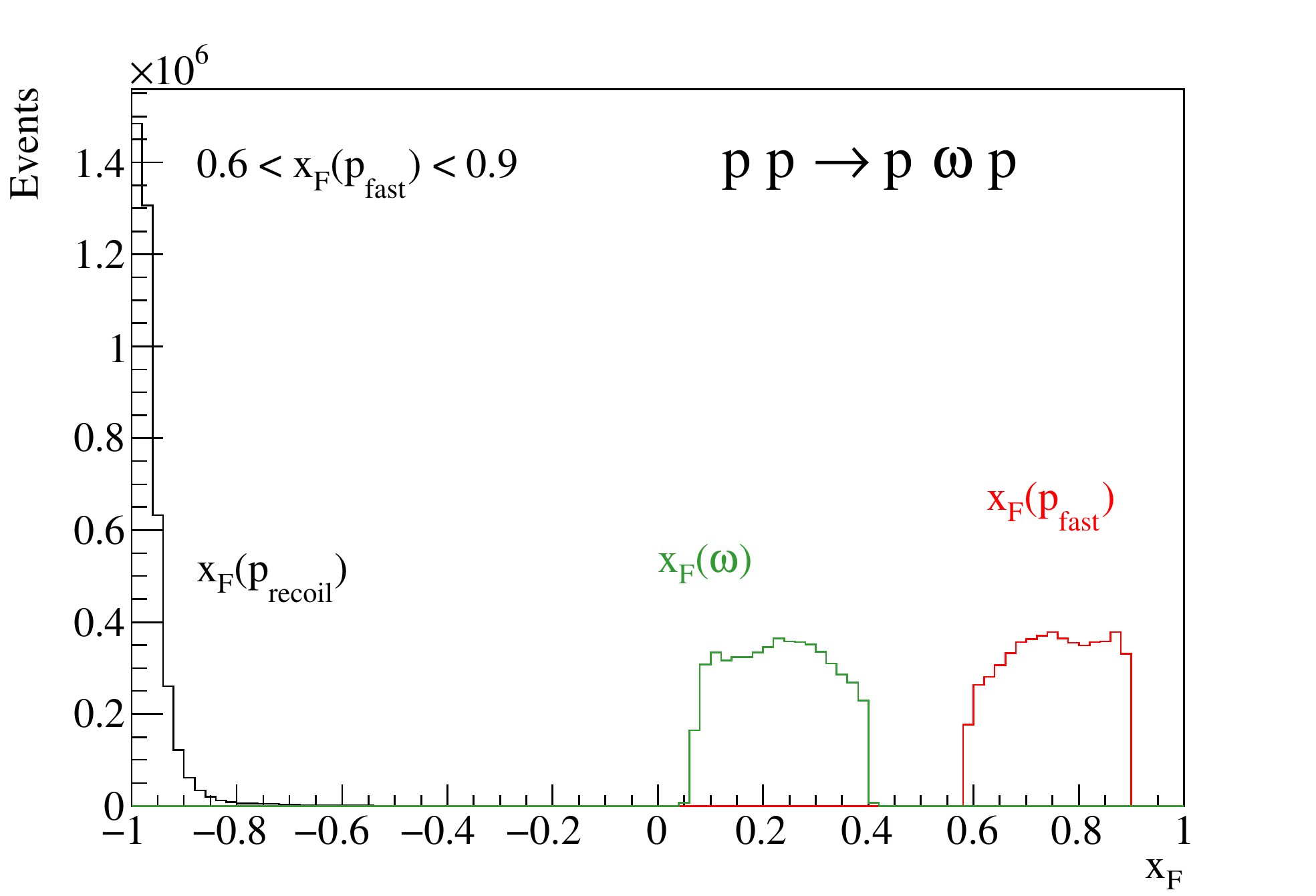}
  \includegraphics[width=.45\textwidth]{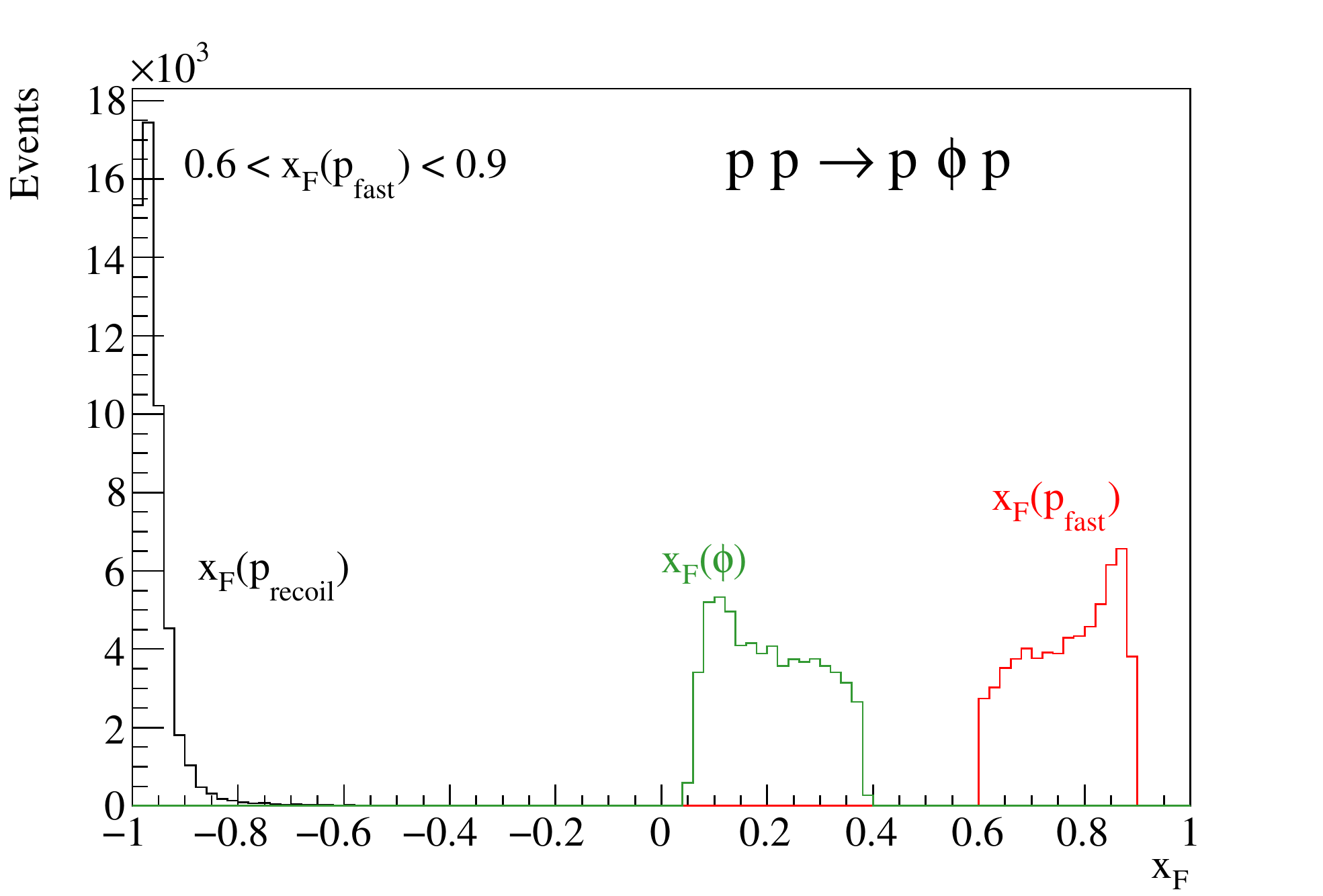}
  \caption[acceptances]{Upper panels: One-dimensional (integrated) acceptances
    for $p\,p\,\rightarrow\,p\,p\,\omega, \omega\,\rightarrow\,\pi^+\pi^-\pi^0$
    (left) and $p\,p\,\rightarrow\,p\,p\,\phi, \phi\,\rightarrow\,K^+K^-$
    (right) as a function of $x_\mathrm{F}$ of the fast proton. Cuts used in the
    later analysis are illustrated by the vertical lines. Middle panels:
    $x_\mathrm{F}$ distributions for
    $pp\,\rightarrow\,pp\,\omega,\,\omega\,\rightarrow\,\pi^+\pi^-\pi^0$ (left)
    and $pp\,\rightarrow\,pp\,\phi,\,\phi\,\rightarrow\,K^+K^-$, acceptance
    uncorrected. Lower panels: The same as shown in the middle panels, but
    acceptance corrected and for $0.6<x_\mathrm{F}<0.9$.} 
\label{fig:acc_xf}
\end{figure}

\subsection{Background subtraction}
\label{sec:bgsub}
The yield of $\phi$ mesons is determined from a fit of a Breit-Wigner shape with
fixed width taken from Ref. \cite{pdg}, which is convoluted with a Gaussian on
top of a background parameterisation that includes $KK$ threshold effects. We
observe a better fit quality using the simple Breit-Wigner functional form
instead of also taking into account $L$-dependent centrifugal barrier terms. All
results in this work are therefore obtained using the simpler Breit-Wigner
function. The used background distribution function is $a\,(m_{K\bar{K}} -
m_1)^n\,(m_{K\bar{K}} - m_2)^k,$ where $a, m_1, m_2, n$\,and\,$k$ are the fit
parameters.

 The yield of $\omega$ mesons is determined from a fit of a Breit-Wigner shape
 as explained above, but this time convoluted with two Gaussians to account for
 different resolutions of the two electromagnetic calorimeters. This fit also
 includes a second-degree polynomial background. Examples of mass spectra for
 the $0.6 < x_\mathrm{F} < 0.7$ region are shown in
 Fig.\,\ref{fig:im_omega_phi}.
 
The sideband subtraction is also used in order to estimate the systematics of
the background subtraction. To obtain background corrected distribution of
\textit{e.g.} $M_{p\mathrm{V}}$, events within $\pm3\sigma$ of the
$M_{\pi^+\pi^-\pi^0}$ or $M_{K^+K^-}$ distributions are taken and events in the
sidebands from $\pm4\sigma$ to $\pm7\sigma$, respectively, are subtracted. The
systematic uncertainty from the background subtraction is estimated by comparing
the yields obtained using different parameterisations of peak and
background. The relative difference of the yields is found to be always below
5\%.

\begin{figure}[h!]
  \centering
  \includegraphics[width=.45\textwidth]{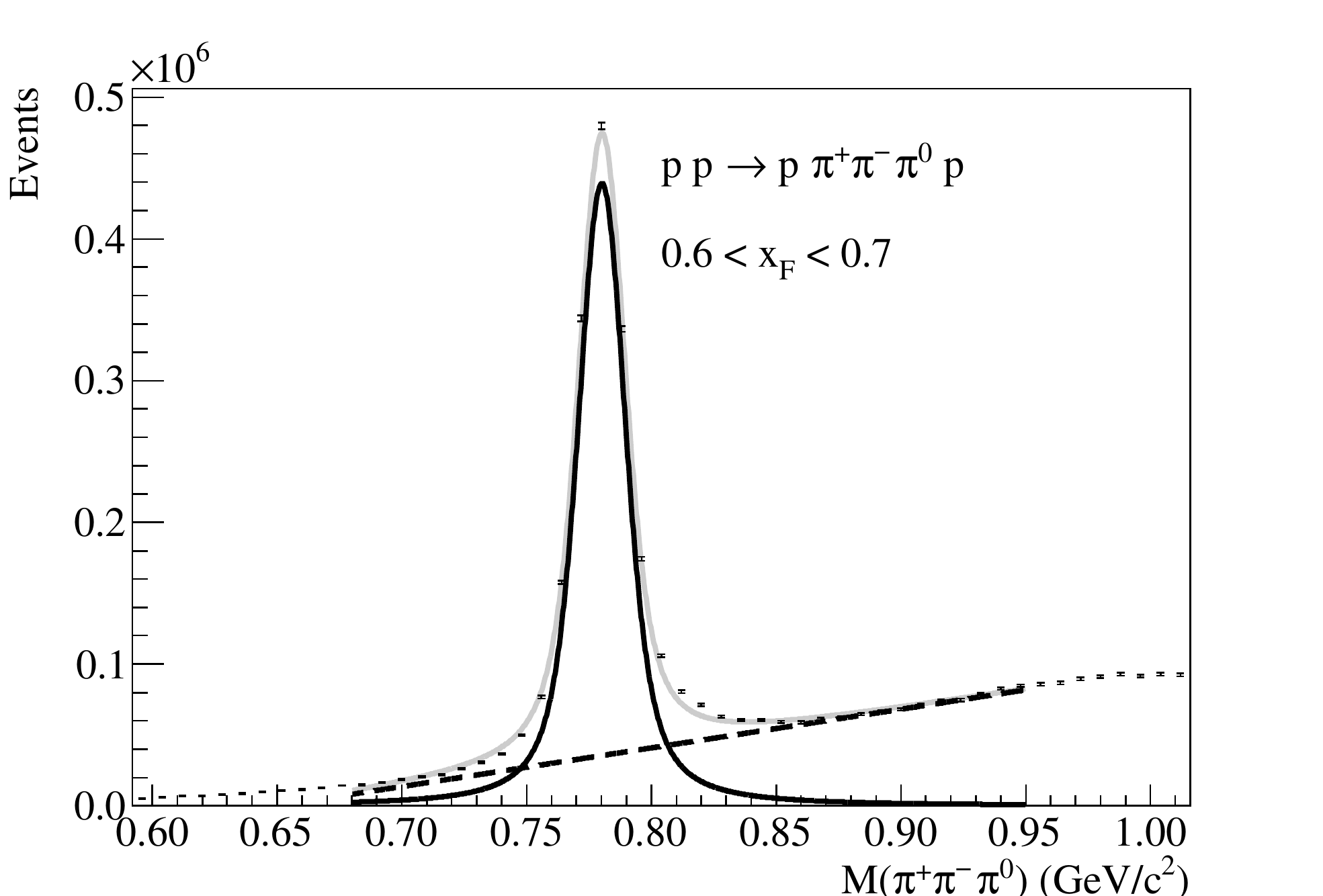}
  \includegraphics[width=.45\textwidth]{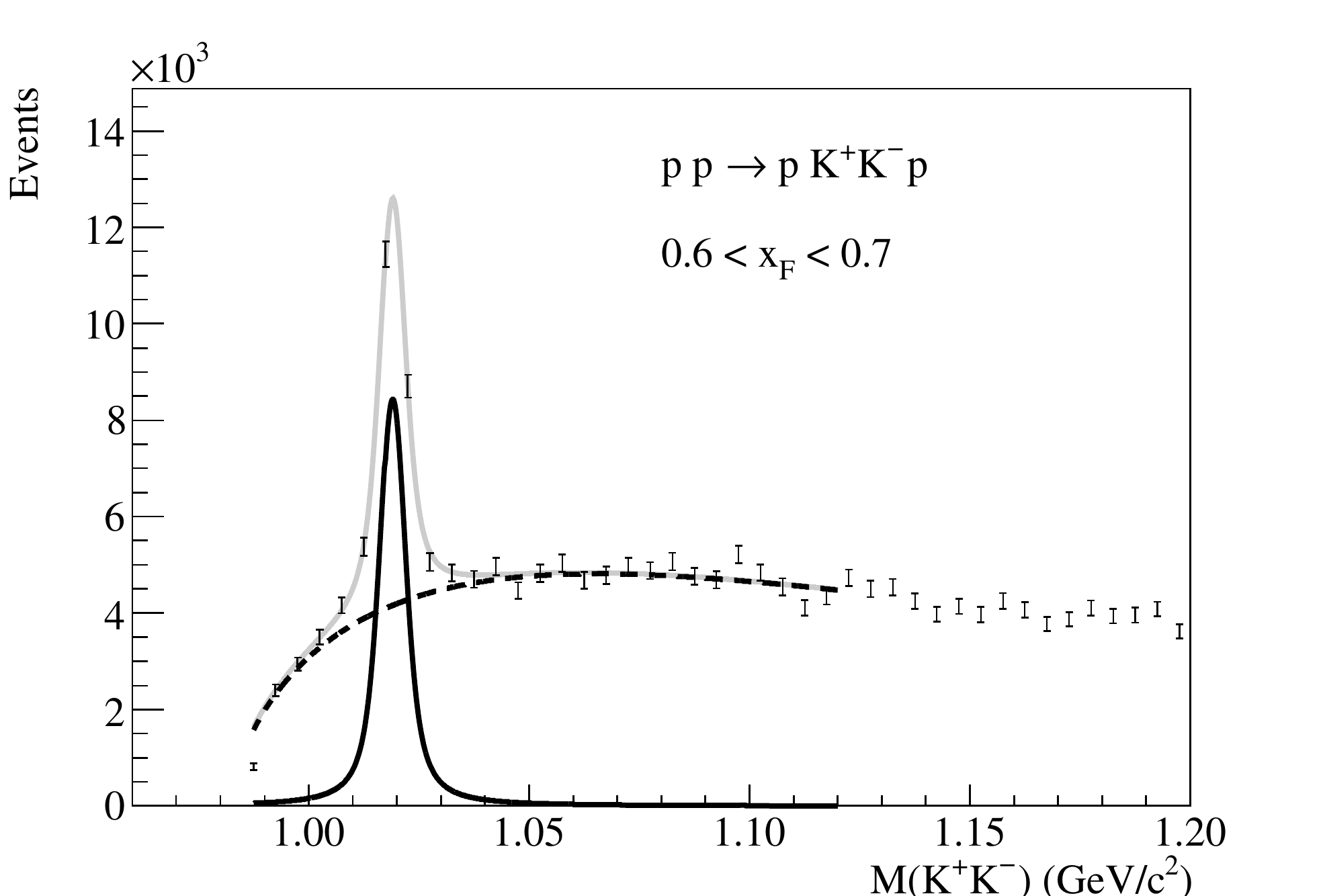}
  \caption[$M_{\pi^+\pi^-\pi^0}$ and $M_{K^+K^-}$]{Left: The fitted mass
    distribution of the $\pi^+\pi^-\pi^0$ system where the $x_\mathrm{F}$ of the
    fast proton is within the interval $0.6<x_\mathrm{F}<0.7$. Right: The fitted
    mass distribution of the $K^+K^-$ system in the $0.6<x_\mathrm{F}<0.7$
    range. The signal fit is shown in black, the background is shown by the
    dashed curve and their sum is shown in grey.} 
  \label{fig:im_omega_phi}
\end{figure}

\subsection{Systematic uncertainties}
\label{sec:sys}
In addition to the uncertainty of the background subtraction, there are other
effects which contribute to the overall systematic uncertainties. Most
efficiencies (CEDAR, RPD, track reconstruction) cancel in
$R_{\phi/\omega}$. Systematic effects introduced by the MC generator are
negligible since a multi-dimensional acceptance correction is applied (see
section \ref{sec:acceptance}). The uncertainty from the RICH is estimated to be
5\% on $R_{\phi/\omega}$ and dominantly stems from background subtraction
uncertainties in the RICH efficiency determination.  The photon reconstruction
efficiency of the ECALs is determined by comparing $\omega$ decays into
$\pi^+\pi^-\pi^0$ and $\pi^0\gamma$ in both real data and MC data with the
assumption that the $\pi^0$ efficiency is the same in both channels. The
deviation between measured efficiency and MC efficiency is found to be below
$10\%$ and used as an upper limit for the systematic uncertainty arising from
the ECALs. The quadratic sum of the 5\% uncertainty from the background
subtraction, the 5\% from the RICH efficiency and the 10\% from the photon
reconstruction efficiency results in a total systematic uncertainty of 12\% for
the results on the cross section ratio quoted in Sec.\,\ref{sec:ratio}.

Uncertainties due to RICH and ECAL efficiencies have no impact on the shape of
angular distributions (Sec.\,\ref{sec:spin}) and $M_{p\mathrm{V}}$ distributions
and thus are neglected. Hence, only the 5\% uncertainty due to background
subtraction is relevant.

\section{$M_{p\mathrm{V}}$ distributions and cross section ratio $R_{\phi/\omega}$}
\subsection{Mass $M_{p\mathrm{V}}$ of the system of fast proton and vector meson}
\label{sec:masspv}
The acceptance-corrected invariant mass distributions of the $pV$ system are
shown in Fig.\,\ref{fig:IM_pphi}. In the case of $\omega$, where the background
is small compared to the signal (see Fig.\,\ref{fig:im_omega_phi}) and has a
locally linear behaviour near the $\omega$ peak, the distributions are obtained
using a sideband subtraction as explained in Sec.\,\ref{sec:sys}. In the
$M_{p\omega}$ spectrum shown to the left in Fig.\,\ref{fig:IM_pphi} several
structures on top of a smooth continuum are clearly discernible. After dividing
the $\omega$ data into finer bins in $x_\mathrm{F}$, as in
Fig.\,\ref{fig:IM_pomega3}, the structures appear even clearer. In the absence
of a partial wave analysis, which is beyond the scope of this paper, the bumps
are compared with known $N^{*}$ resonances. The high-mass bumps are consistent
with resonances listed in the PDG\,\cite{pdg}: the one at about 2.2\,GeV/$c^2$
with $N^{*}(2190)$ $J^{P}=\frac{7}{2}^{-}$, $N^{*}(2200)$
$J^{P}=\frac{9}{2}^{+}$ and $N^{*}(2250)$ $J^{P}=\frac{9}{2}^{-}$ and the one at
about 2.6\,GeV/$c^2$ with $N^{*}(2600)$ $J^{P}=\frac{11}{2}^{-}$ and
$N^{*}(2700)$ $J^{P}=\frac{13}{2}^{+}$. These prominent resonances have high
spin.

The $p\phi$ mass spectrum (Fig.\,\ref{fig:IM_pphi}, right panel) is obtained
using a fit for background subtraction, as explained in
Section\,\ref{sec:bgsub}. It appears without pronounced structures, also
consistent with earlier findings\,\cite{pdg}.

\begin{figure}[h!]
  \centering
  \includegraphics[width=.45\textwidth]{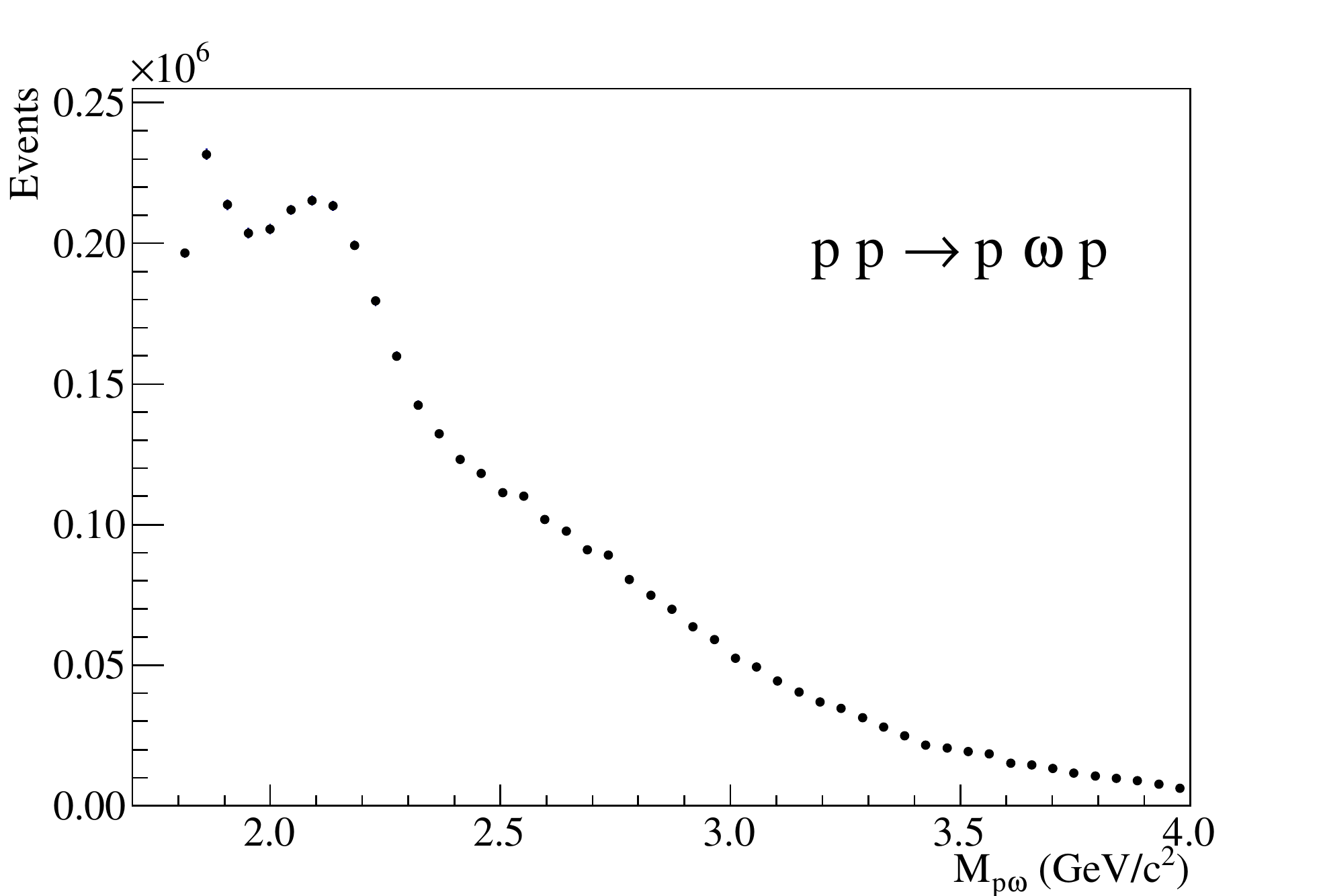}
  \includegraphics[width=.45\textwidth]{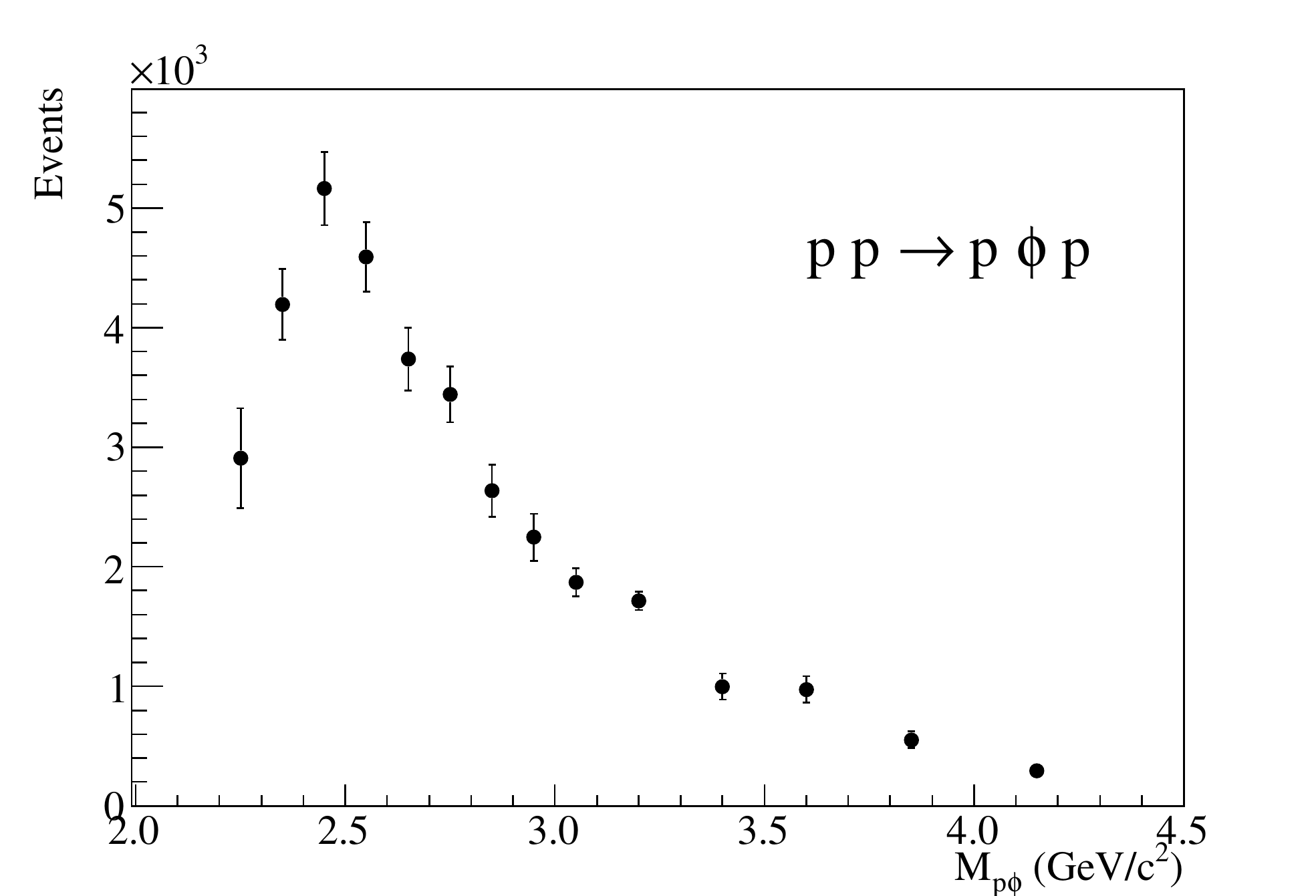}
  \caption[Mass distributions]{Distributions of the invariant mass of the $pV$
    system for $0.6 < x_\mathrm{F} < 0.9$. Left: The $M_{p\omega}$ spectrum. The
    background is subtracted using the sideband method. Right: The $M_{p\phi}$
    spectrum. The background is subtracted using a polynomial fit described in
    Section\,\ref{sec:bgsub} and the uncertainty from the fit is included in the
    error bars.} 
  \label{fig:IM_pphi}
\end{figure}

\begin{figure}[h!]
  \centering
  \includegraphics[width=.45\textwidth]{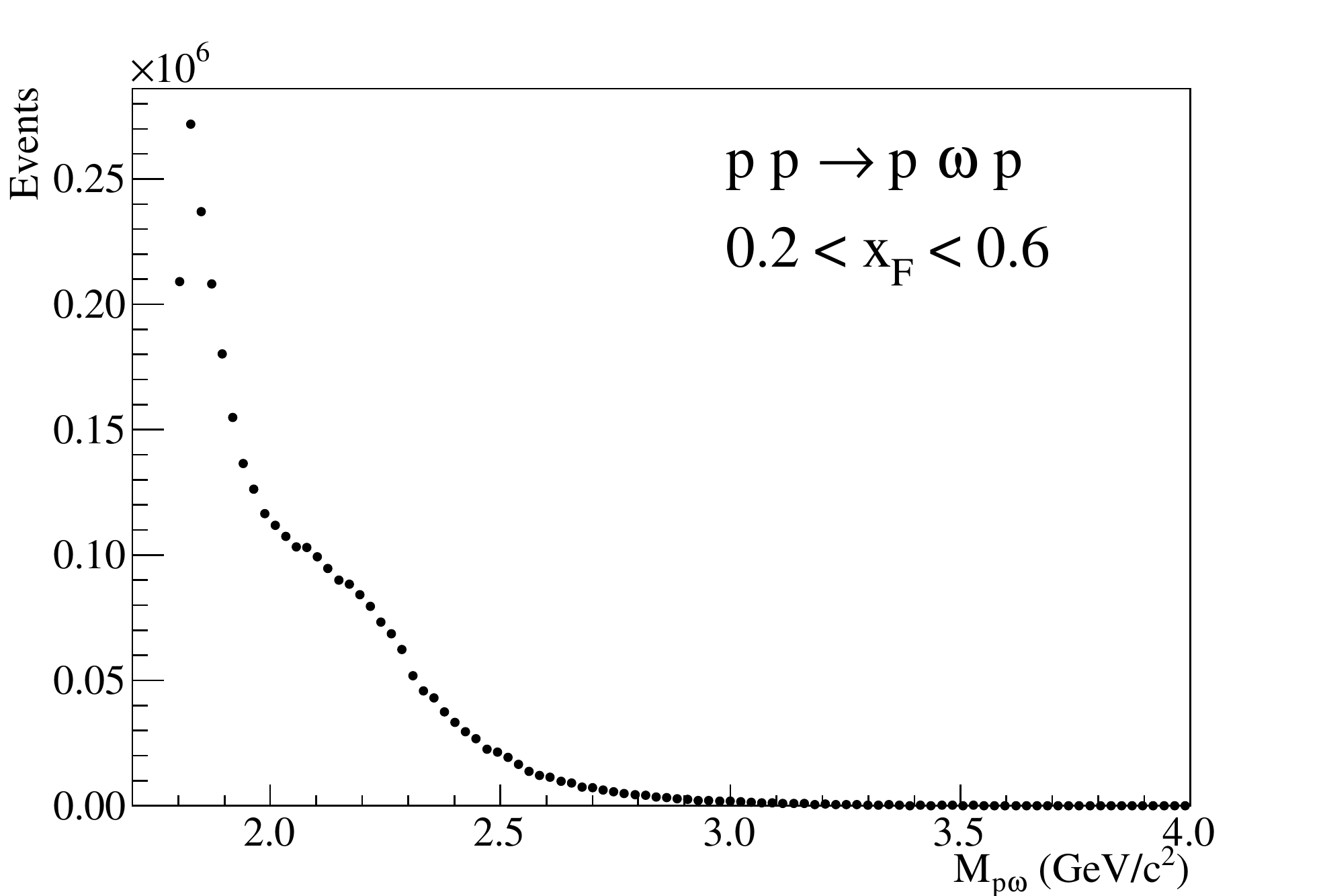}
  \includegraphics[width=.45\textwidth]{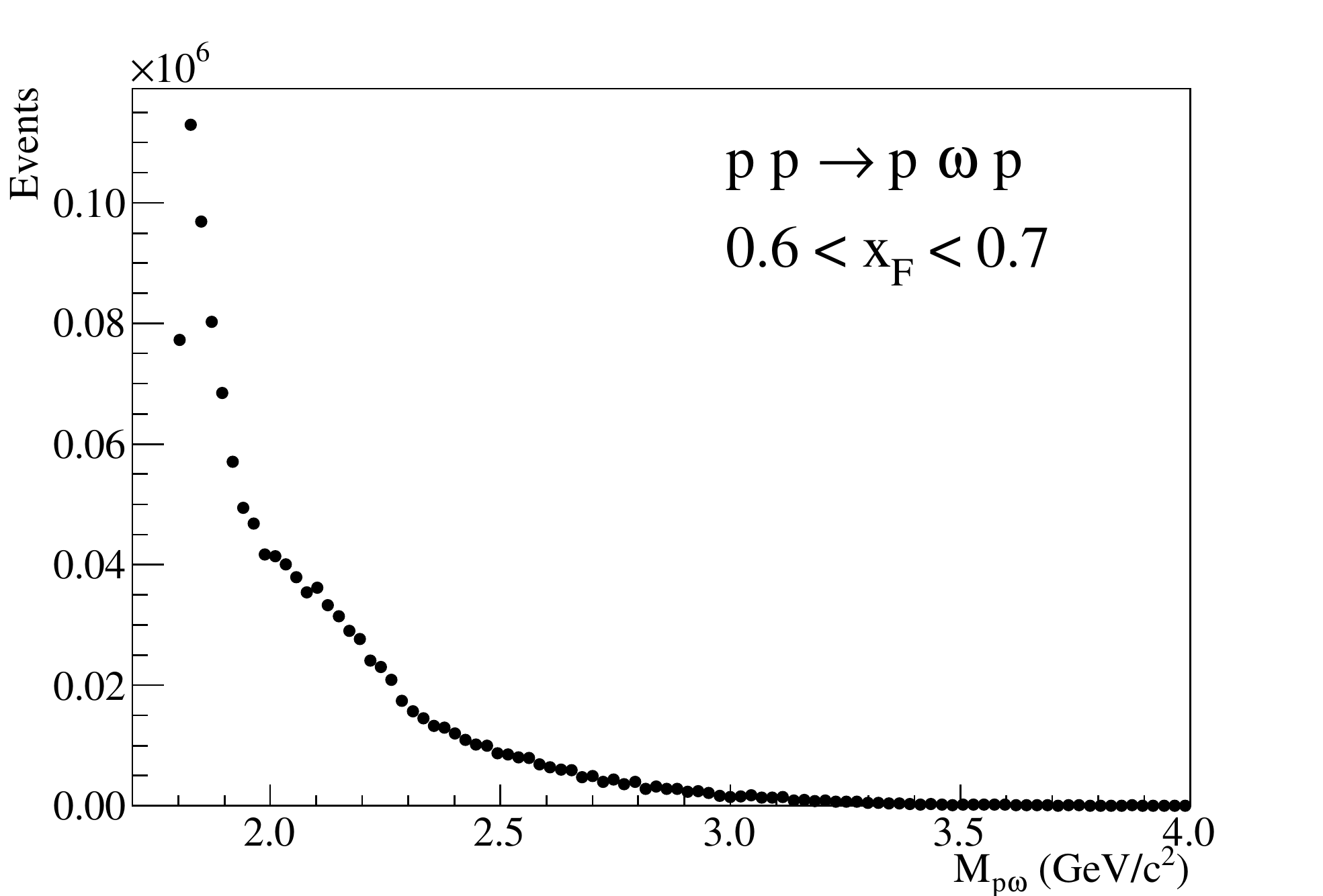}\\
  \includegraphics[width=.45\textwidth]{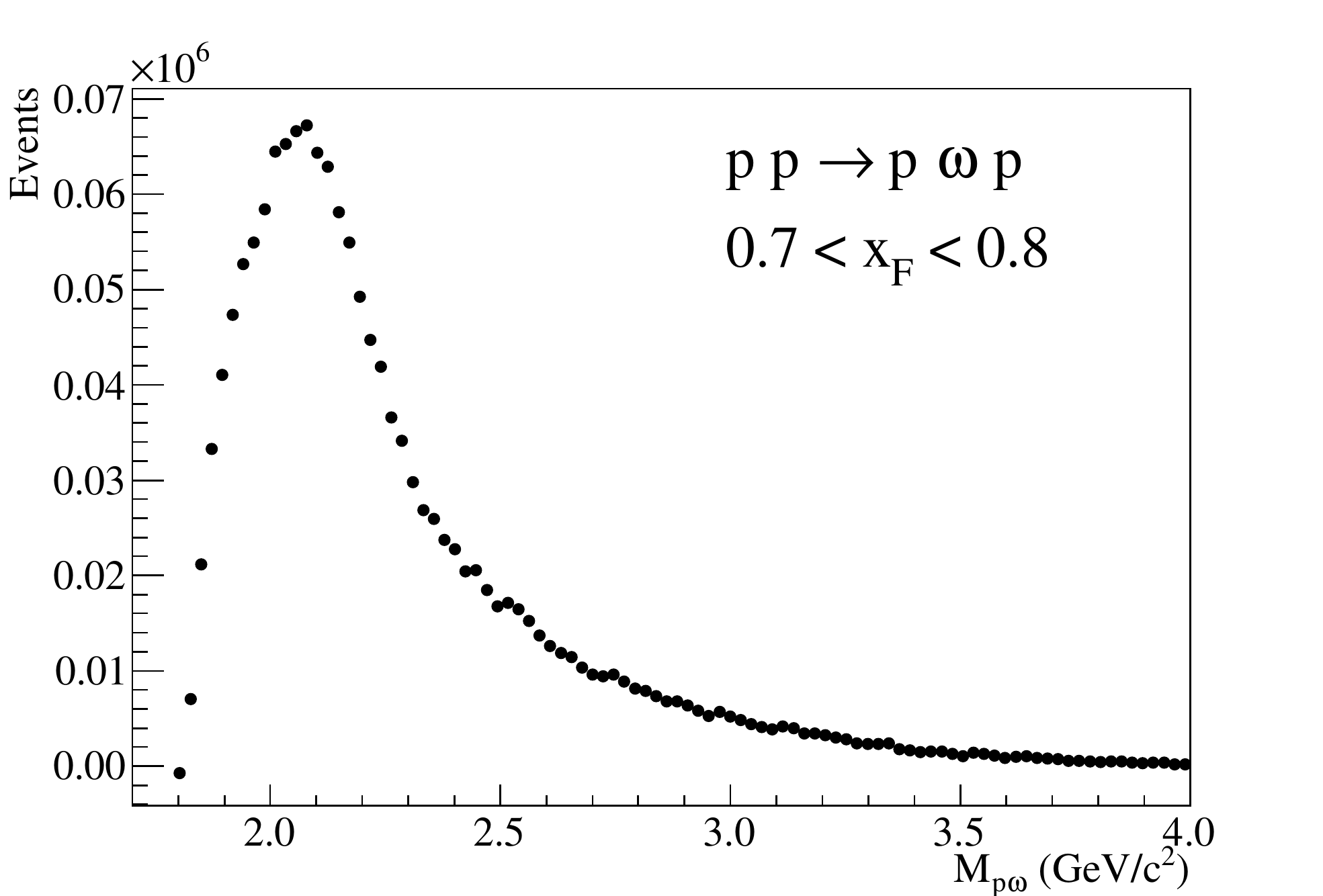}
  \includegraphics[width=.45\textwidth]{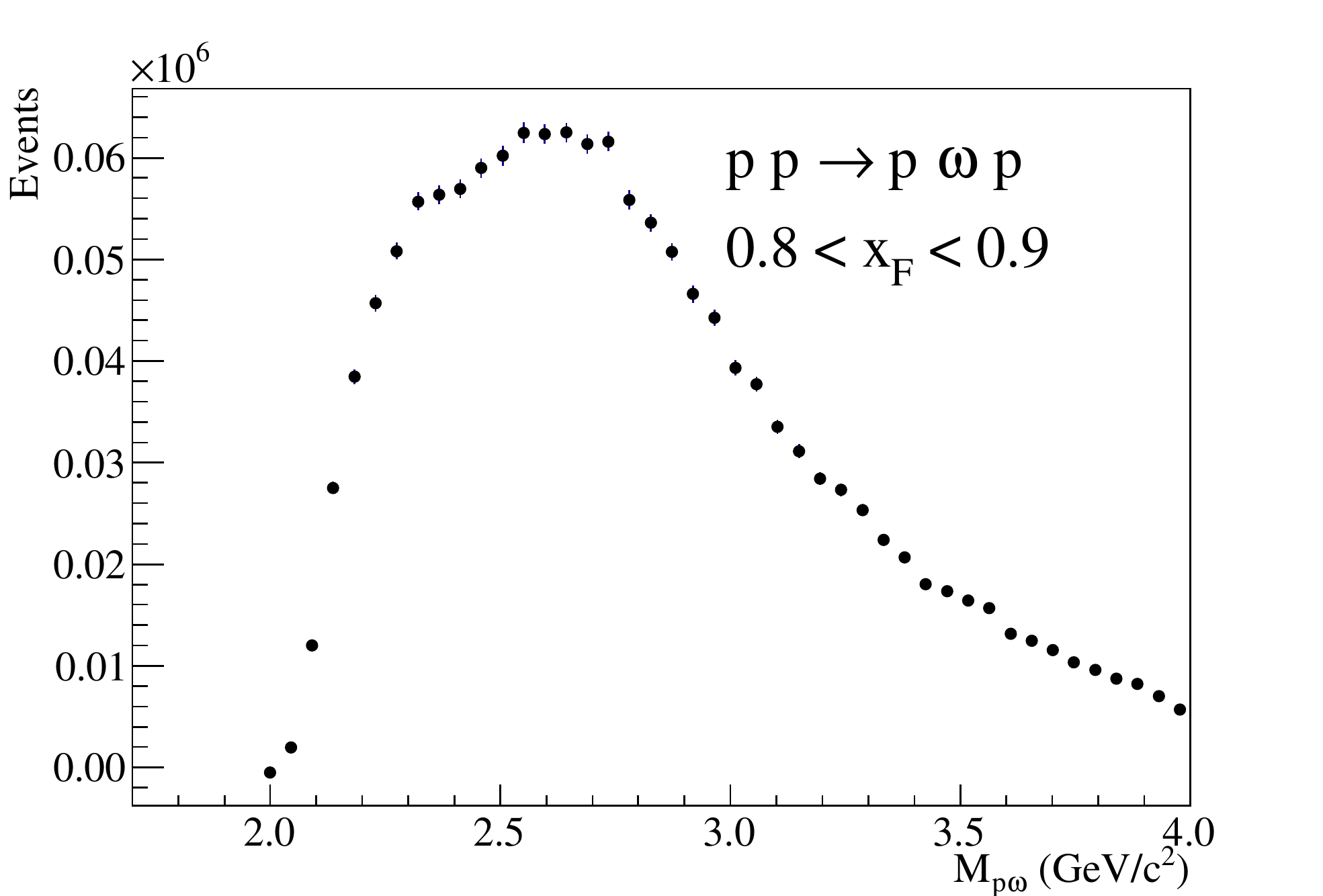}
  \caption[$M_{p\omega}$]{Distributions of the mass of the $p$-$\omega$ system
    for $0.2 < x_\mathrm{F} < 0.6$ (upper left), $0.6< x_\mathrm{F} < 0.7$
    (upper right), $0.7< x_\mathrm{F} < 0.8$ (lower left) and $0.8< x_\mathrm{F}
    < 0.9$  (lower right).} 
  \label{fig:IM_pomega3}
\end{figure}

\subsection{Cross section ratio $R_{\phi/\omega}$}
\label{sec:ratio}
The $\pi^+\pi^-\pi^0$ and $K^+K^-$ data are divided into three intervals of
$x_\mathrm{F}$: 0.6--0.7, 0.7--0.8 and 0.8--0.9. In each interval, the
acceptance-corrected $\omega$ and $\phi$ yields are calculated using the method
described in Sec.\,\ref{sec:bgsub} and corrected for the branching ratios of the
$\omega \rightarrow \pi^+\pi^-\pi^0$ and $\phi \rightarrow K^+K^-$ decays,
respectively. The ratio $R_{\phi/\omega}$ is calculated in each $x_\mathrm{F}$
interval. The results, summarised in Table\,\ref{tab:resultsxF_combined} and
Fig.\,\ref{fig:resultsxF_combined}, show that the OZI rule is violated by a
factor $F_\mathrm{OZI}$ of 4.5, 4.0 and 2.9, \textit{i.e.} $\phi$ production is
enhanced with respect to the OZI rule prediction. The violation factor is
defined as $F_\mathrm{OZI}=R_{\phi/\omega}/\tan^2\delta_\mathrm{V}$, with
$\tan^2\delta_\mathrm{V}=0.0042$ being the OZI prediction. It is notable that
the violation is smaller in the highest $x_\mathrm{F}$ bin.  
The average value $\langle R
\rangle_{\phi/\omega}\,=\,0.0160\,\pm\,0.0003\,\pm\,0.0020$ is consistent with
the result from SPHINX\,\cite{sphinx1}, which is $\langle R
\rangle_{\phi/\omega}\,=\,0.0155\,\pm\,0.0005\,\pm\,0.0031$. 

\begin{table}[h!]
\caption{Differential cross section ratios
  $R_{\phi/\omega}\,=\,\frac{d\sigma(p\,p\,\rightarrow\,p\,\phi\,p)/dx_\mathrm{F}}{d\sigma(p\,p\,\rightarrow\,p\,\omega\,p)/dx_\mathrm{F}}$
  and corresponding OZI violation factors $F_\mathrm{OZI}$.}
\begin{center}

        \begin{tabular}[htp]{cccccc}
          \hline
                $x_\mathrm{F}$ & $R_{\phi/\omega}$ & Stat. & Fit & Syst. & $F_\mathrm{OZI}$ \\
                \hline
                0.6--0.7 & 0.019 & 0.0003 & 0.0006 & 0.0023 & $4.5 \pm 0.6$ \\
		0.7--0.8 & 0.017 & 0.0002 & 0.0004 & 0.002 & $4.0 \pm 0.5$ \\
		0.8--0.9 & 0.012 & 0.0002 & 0.0005 & 0.0014 & $2.9 \pm 0.4$\\
        \end{tabular}
\end{center}
	
	\label{tab:resultsxF_combined}
\end{table}
The $M_{p\omega}$ distributions shown in Fig.\,\ref{fig:IM_pomega3} indicate
that the $p\,p\,\rightarrow\,p\,\omega\,p$ cross section may be heavily
influenced by the baryon resonances. Unless the resonant contribution is removed
from the data set, a measurement of the cross section ratio $R_{\phi/\omega}$
does not give sufficient information, neither about the strangeness content of
the nucleon nor about other production mechanisms than resonant diffractive
production. No resonances are visible above $M_{p\omega}$ = 3.3\,GeV/$c^2$. For
a consistent treatment of $\phi$ and $\omega$ production, the vector meson
momentum $p_\mathrm{V}$ is used as determined in the $pV$ rest system: 
\begin{equation}
p_\mathrm{V}\,=\,\frac{\sqrt{\left( M_{p\mathrm{V}}^2\,-\,(m_\mathrm{V}\,+\,m_p)^2 \right)\left(M_{p\mathrm{V}}^2\,-\,(m_\mathrm{V}\,-\,m_p)^2 \right)}}{2\,M_{p\mathrm{V}}}.
\label{eq:p_V}
\end{equation}
\noindent The mass value $M_{p\omega}$ = 3.3\,GeV/$c^2$ corresponds to
$p_\mathrm{V} = $1.4\,GeV$/c$, which is hence used as a cut value also for the
$\phi$ meson. The requirement of $p_\mathrm{V} >$ 1.4\,GeV$/c$ results in ratios
of 0.034 and 0.032 in the two bins $0.7<x_\mathrm{F}<0.8$ and
$0.8<x_\mathrm{F}<0.9$, respectively, which correspond to OZI violation factors
$F_\mathrm{OZI}=7.9$ and $F_\mathrm{OZI}=7.6$. In the bin
$0.6<x_\mathrm{F}<0.7$, the $\phi$ yield is insufficient for a reliable
$R_{\phi/\omega}$ estimate. Detailed results are summarised in the bottom part
of Table\,\ref{tab:resultsxFp} and in Fig.\,\ref{fig:resultsxF_combined}.  

Note that if the low-mass resonant region in $M_{p\omega}$ is removed, this
results in an OZI violation factor of about 8, independent of $x_\mathrm{F}$ in
the observed range. This agrees well with the results from the SPHINX experiment
that operated at a beam energy of 70\,GeV\,\cite{sphinx1}. In order to remove
the resonant region, SPHINX applied a weaker cut of 1\,GeV$/c$ on the
$p_\mathrm{V}$ momentum. This corresponds to mass values of $M_{p\omega}$ of
2.64\,GeV/$c^2$  and $M_{p\phi}$ of 2.8\,GeV/$c^2$. Applying the same cut on the
COMPASS data gives ratios $R_{\phi/\omega}=0.032$, 0.038 and 0.019 in the three
$x_\mathrm{F}$ bins, which correspond to OZI violation factors
$F_\mathrm{OZI}=7.6$, 9 and 4.5 respectively, as summarised in the top part of
Table\,\ref{tab:resultsxFp} and Fig.\,\ref{fig:resultsxF_combined}. The COMPASS
results below $x_\mathrm{F}=0.8$ are consistent with the SPHINX result
$\frac{\sigma(p\,N\,\rightarrow\,p\,N\,\phi)}{\sigma(p\,N\,\rightarrow\,p\,N\,\omega)}\,=\,0.040\,\pm\,0.0004\,\pm
0.008$. The $x_\mathrm{F}$ range of the SPHINX data is not stated explicitly in
Ref.\,\cite{sphinx1}.    

\begin{table}[h!]
\caption{Differential cross section ratio $R_{\phi/\omega}$ and corresponding
  OZI violation factors $F_\mathrm{OZI}$ for different $p_\mathrm{V}$ cuts.} 
\begin{center}

        \begin{tabular}[htp]{ccccccc}
          \hline
                $p_\mathrm{V}$ (GeV/$c$) & $x_\mathrm{F}$ & $R_{\phi/\omega}$ & Stat. & Fit & Syst. & $F_\mathrm{OZI}$ \\
                \hline
                $>$ 1.0  & 0.6--0.7 & 0.032 & 0.0007 & 0.0013 & 0.0038 & $7.6 \pm 1.0$ \\
		$>$ 1.0  & 0.7--0.8 & 0.038 & 0.0006 & 0.0010 & 0.0046 & $9.0 \pm 1.1$ \\
		$>$ 1.0  & 0.8--0.9 & 0.019 & 0.0003 & 0.0005 & 0.0023 & $4.5 \pm 0.6$ \\
\hline
		$>$ 1.4 & 0.7--0.8 & 0.033 & 0.0013 & 0.0025 & 0.0040 & $7.9 \pm 1.1$ \\
		$>$ 1.4 & 0.8--0.9 & 0.032 & 0.0011 & 0.0017 & 0.0038 & $7.6 \pm 1.0$ \\
        \end{tabular}
\end{center}
	
	\label{tab:resultsxFp}
\end{table}

\begin{figure}[h!]
  \centering
 \includegraphics[width=0.5\textwidth]{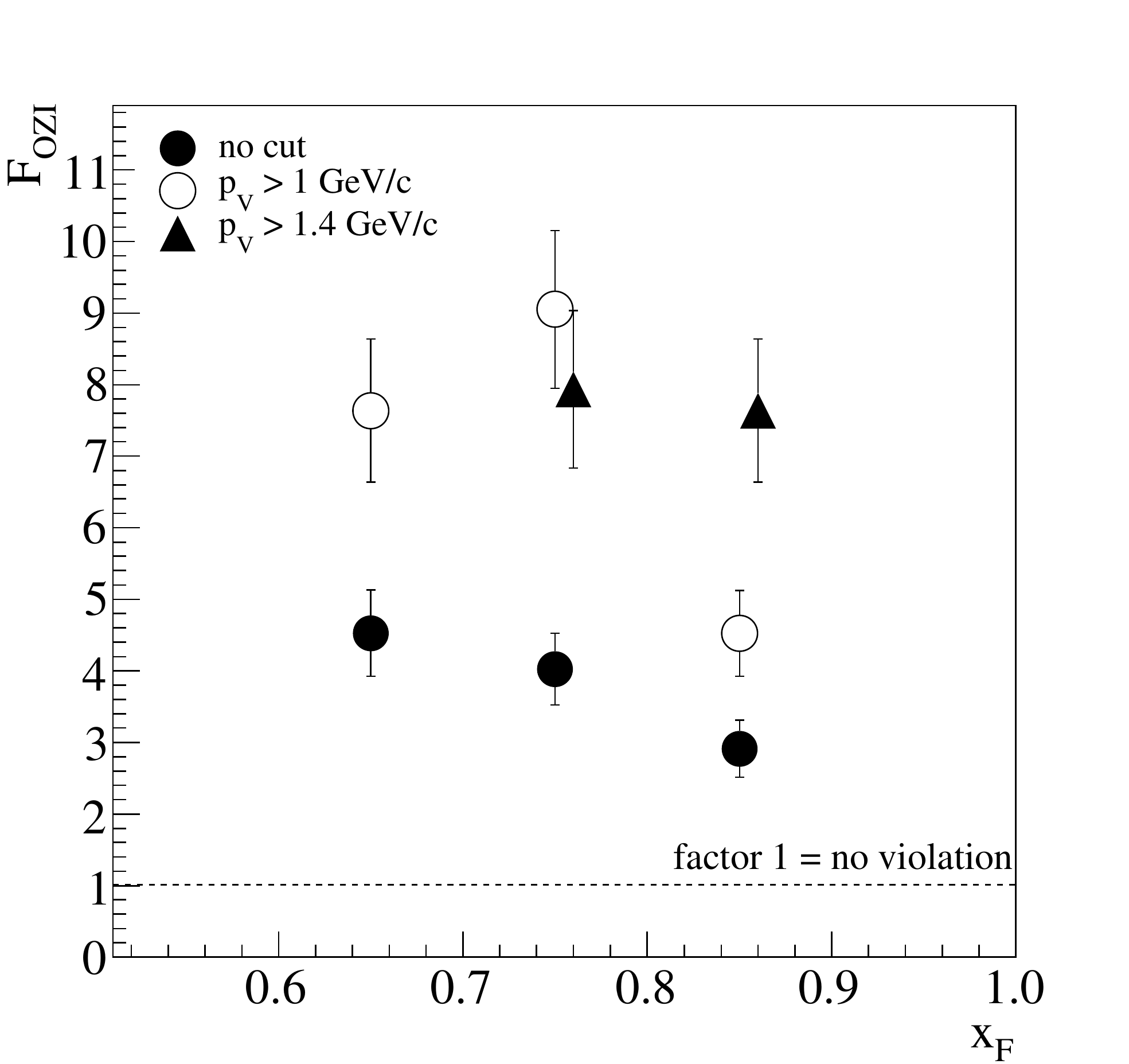}
 
 \caption[$R_{\phi/\omega}$]{OZI violation factor $F_\mathrm{OZI}$ as a function of $x_\mathrm{F}$ for different $p_V$ cuts.}
 \label{fig:resultsxF_combined}
 \end{figure}

\section{Results on spin alignment}
\label{sec:spin}
In order to get more information about production mechanisms, in particular to
find out whether they are the same or different for $\omega$ and $\phi$, it is
helpful to study the spin-alignment (tensor polarisation) of the produced vector
mesons with respect to a given quantisation axis. For different production
processes, the preferential axis of alignment of the vector meson may be
different. In this section, we study the spin alignment by determining the
distributions of the angle between the analyser, defined by the direction of the
decay particles of the vector meson, and two different quantisation axes. 

In the 3-body decay of the $\omega$ meson, the normal to the decay plane is the
most sensitive analyser\,\cite{chung}. In the case of a vector meson decaying
into two pseudoscalars, \textit{e.g.} $\phi \rightarrow K^+K^-$, one chooses the
momentum vector of either one. Schilling, Seyboth and Wolf\,\cite{schilling}
describe the strong decay of a spin-one particle into either two or three
pseudoscalars in terms of the spin-density matrix $\rho$ and the decay matrix
$T$, obtaining the following angular distribution: 
\begin{align}
W\left(\cos\theta,\phi\right) & =  \Tr\{T^*\rho T\} \nonumber \\ 
& = \frac{3}{8\pi} \left( \rho_{11}\sin^2\theta\,+\,\rho_{00}\cos^2\theta-\sqrt{2}\rho_{10}\sin2\theta\cos\varphi-\rho_{1-1}\sin^2\theta\cos2\varphi \right).
\label{eq:fullang}
\end{align}
\noindent Integrating over the azimuthal angle $\varphi$, and using $\Tr\{\rho\}
= 1 = \rho_{00}+\rho_{11}+\rho_{-1-1}$ combined with the symmetry requirement
$\rho_{11}=\rho_{-1-1}$ simplifies Eq.\,(\ref{eq:fullang}) to: 
\begin{equation}
W\left(\cos\theta\right)\,=\frac{3}{4}\left(\,1\,-\,\rho_{00}\,+\,(3\,\rho_{00}\,-\,1)\,\cos^2\theta\right).
\label{eq:2body}
\end{equation}
\noindent For $\rho_{00} = 1/3$, one obtains isotropic angular distributions. 
If $\rho_{00} = 0$, we have a $\sin^2\theta$ dependence and the vector mesons are in the magnetic sub-state $M=\pm 1$ with respect to the quantisation axis, while $\rho_{00} = 1$ gives a pure $\cos^2\theta$ dependence and corresponds to $M=0$.

In the figures of this section, the error bars represent the quadratic sum of
statistical uncertainty and the point-to-point uncertainty of the background
subtraction.   
\subsection{Spin alignment with respect to the direction of the $pV$ system}
\label{sec:spinhel}
The spin alignment is first studied in the $pV$ helicity frame. The reference
axis ($z$-axis) is the direction of the $pV$ system in the rest system of the
vector meson $V$. If, on the one hand, the vector meson results from a
diffractively produced baryon resonance, the spin alignment of the vector meson
is expected to be sensitive to the direction of this resonance. If on the other
hand the dominating process is a central Reggeon--Reggeon/Reggeon--Pomeron
fusion or in the absence of a resonant system, there is no longer a preferred
reference axis and the distributions are expected to be isotropic.  The polar
angle of an analyser in the helicity frame will in the following be referred to
as ``helicity angle'' and be denoted by $\theta_H$. The $\cos^2\theta_H$
distributions are shown in Fig.\,\ref{fig:cos2theta_xf} in different
$x_\mathrm{F}$ intervals. The background distribution (open circles) is obtained
by sideband subtraction and found to be isotropic. A striking feature of the
signal data is that the slope is varying with $x_\mathrm{F}$ in the case of the
$\omega$ meson (see Fig.\,\ref{fig:cos2theta_xf}, left), going from a strong
negative slope in the interval $0.2<x_\mathrm{F}<0.6$ passing through isotropy
in the interval $0.7<x_\mathrm{F}<0.8$ to a strong positive slope in the
interval $0.8<x_\mathrm{F}<0.9$. No such behaviour is observed in the case of
the $\phi$ meson (see Fig.\,\ref{fig:cos2theta_xf}, right), for which the
distributions are fairly isotropic in all three $x_\mathrm{F}$ intervals between
0.6 and 0.9. In the case of the $\phi$ meson, it should however be pointed out
that the statistical uncertainty is significantly larger compared to the case of
$\omega$ and it is difficult to draw definite conclusions from the $\phi$ decay
angular distributions.
\begin{figure}[htp]
  \centering
  \includegraphics[width=\textwidth]{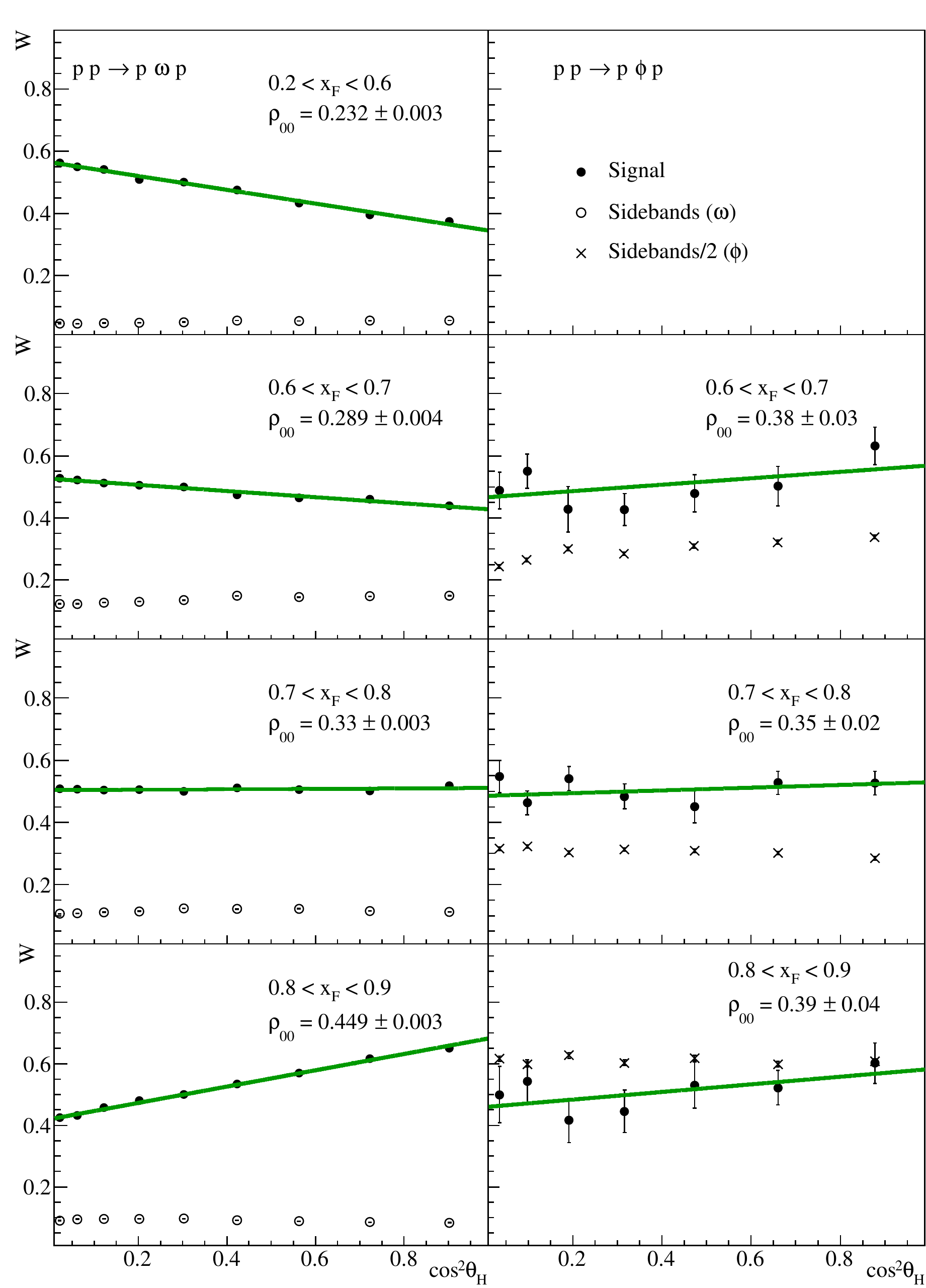}
  \caption[Helicity angle distributions]{The closed points represent the angular
    distributions of $\cos^{2}\theta$, where $\theta = \theta_H$ is the helicity
    angle of the $\omega$ meson (right panels) and of the $\phi$ meson (left
    panels) in different $x_\mathrm{F}$ regions. The open points show the
    corresponding distribution for the events in the sidebands around the
    $\omega$ peak in the $M(\pi^+\pi^-\pi^0)$ distribution. The crosses show the
    corresponding distribution (scaled by 0.5) for the events in the sidebands
    around the $\phi$ peak in the $M(K^+K^-)$ distribution. The lines are the
    results of linear fits as explained in the text.} 
  \label{fig:cos2theta_xf}
\end{figure}

The $\rho_{00}$ element is extracted by fitting straight lines $a + bx$, $x =
\cos^2\theta_H$ to the data points and then solving Eq.\,\ref{eq:2body}. The
fits were performed with and without including the leftmost and the rightmost
data points in the angular distributions. The difference is included in the
uncertainty. For $\omega$, the contribution to the total uncertainty is very
small. For $\phi$ it is typically between 5\% and 10\%.  The fit results are
shown in Table\,\ref{tab:rho00res} and Fig.\,\ref{fig:rho00res} including those
for $p_{\omega}>$ 1\,GeV/$c$. Within uncertainties, no $\phi$ meson spin
alignment is observed with respect to the $p\phi$ direction. Similarly, the
$\omega$ meson alignment with respect to the $p\omega$ direction almost vanishes
for $p_{\omega}>$ 1\,GeV/$c$ and $x_\mathrm{F}<0.8$. For $p_\omega >
1.4$\,GeV/$c$, above the low-mass resonant region, the angular distribution of
the $\omega$ meson decay is, within the larger uncertainty, consistent with
isotropy even when $x_\mathrm{F} > 0.8$.

\begin{table}[h]
\caption{Spin alignment $\rho_{00}$ extracted from the helicity angle
  distributions for $\phi$ and $\omega$ production, in the latter case with
  various cuts on $p_\omega$. The uncertainty is the propagated uncertainty from
  the linear fits, which in turn includes the quadratic sum of statistical
  uncertainties and uncertainties from the background subtraction.} 
\begin{center}
        \begin{tabular}[htp]{lcc}
          \hline
                Reaction & $x_\mathrm{F}$ & $\rho_{00}$ \\
                \hline
$pp \rightarrow pp \phi, \phi \rightarrow K^+ K^-$ & 0.6--0.7 & $0.38 \pm 0.03$ \\
$pp \rightarrow pp \phi, \phi \rightarrow K^+ K^-$ & 0.7--0.8 & $0.35 \pm 0.02$ \\
$pp \rightarrow pp \phi, \phi \rightarrow K^+ K^-$ & 0.8--0.9 & $0.39 \pm 0.04$ \\
\hline
$pp \rightarrow pp \omega, \omega \rightarrow \pi^+\pi^-\pi^0$ & 0.2--0.6 & $0.232 \pm 0.003$ \\
$pp \rightarrow pp \omega, \omega \rightarrow \pi^+\pi^-\pi^0$ & 0.6--0.7 & $0.289 \pm 0.004$ \\	
$pp \rightarrow pp \omega, \omega \rightarrow \pi^+\pi^-\pi^0$ & 0.7--0.8 & $0.330 \pm 0.003$ \\
$pp \rightarrow pp \omega, \omega \rightarrow \pi^+\pi^-\pi^0$ & 0.8--0.9 & $0.449 \pm 0.003$ \\
\hline
$pp \rightarrow pp \omega, \omega \rightarrow \pi^+\pi^-\pi^0, p_\mathrm{V} > 1.0$ GeV/$c$ & 0.2--0.6 & $0.30 \pm 0.01$ \\
$pp \rightarrow pp \omega, \omega \rightarrow \pi^+\pi^-\pi^0, p_\mathrm{V} > 1.0$ GeV/$c$ & 0.6--0.7 & $0.34 \pm 0.01$ \\	
$pp \rightarrow pp \omega, \omega \rightarrow \pi^+\pi^-\pi^0, p_\mathrm{V} > 1.0$ GeV/$c$ & 0.7--0.8 & $0.306 \pm 0.006$ \\
$pp \rightarrow pp \omega, \omega \rightarrow \pi^+\pi^-\pi^0, p_\mathrm{V} > 1.0$ GeV/$c$ & 0.8--0.9 & $0.463 \pm 0.003$ \\
\hline
$pp \rightarrow pp \omega, \omega \rightarrow \pi^+\pi^-\pi^0, p_\mathrm{V} > 1.4$ GeV/$c$ & 0.8--0.9 & $0.37 \pm 0.03$ \\

        \end{tabular}
\end{center}
	
	\label{tab:rho00res}
\end{table}
\begin{figure}[htp]
  \centering
  \includegraphics[width=0.9\textwidth]{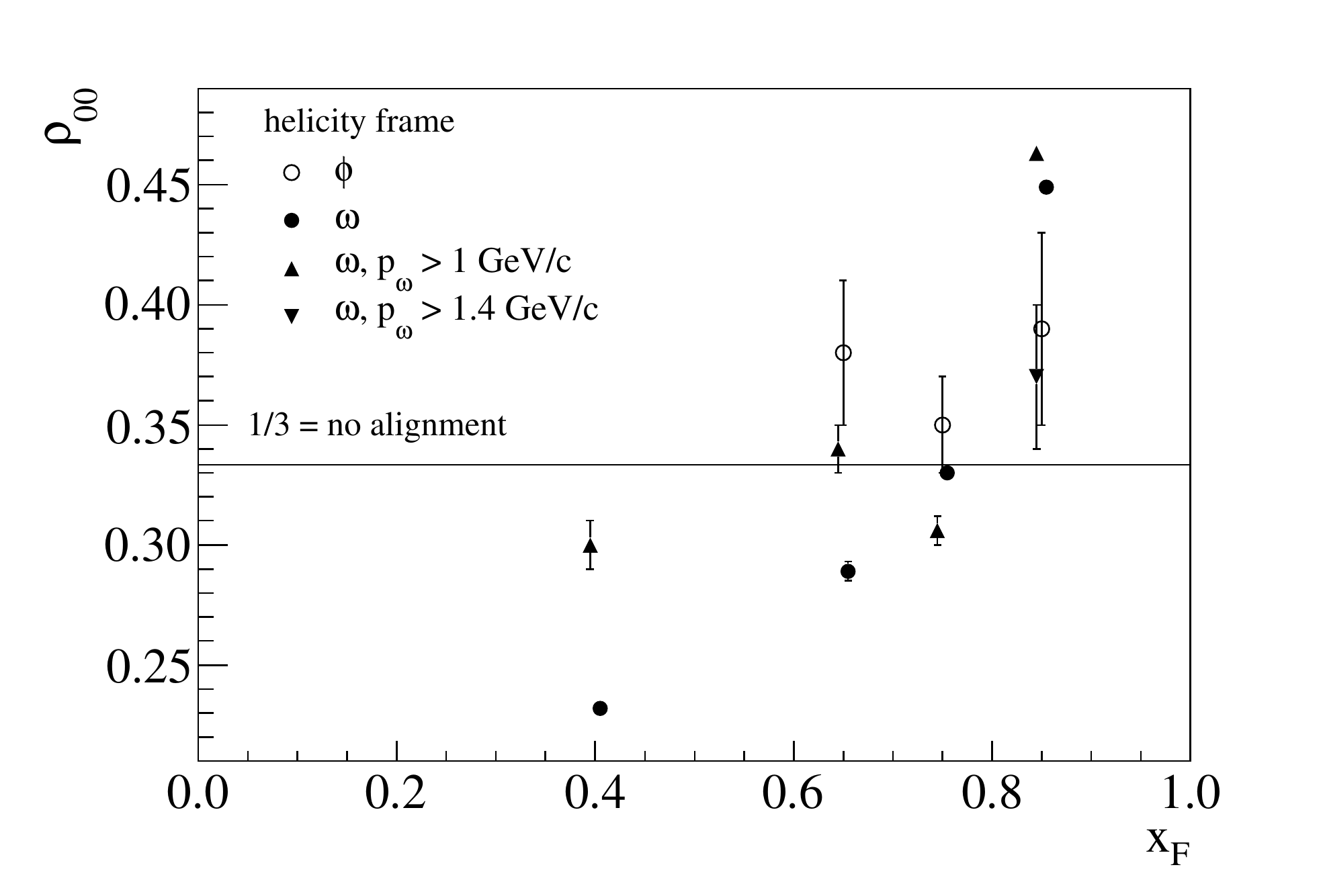}
  \caption[]{Spin alignment $\rho_{00}$ extracted from the helicity angle
    distributions for $\phi$ and $\omega$ production as a function of
    $x_\mathrm{F}$ for several cuts on $p_\mathrm{V}$.} 
 \label{fig:rho00res}
\end{figure}

Extracting helicity angle distributions in slices of $M_{p\omega}$ reveals a
clear dependence of $\rho_{00}$ on $M_{p\omega}$, see
Figs.\,\ref{fig:cos2theta_Mpomega} and \ref{fig:rho00M} and
Table\,\ref{tab:rho00M}. The dependence of $\rho_{00}$ on $x_\mathrm{F}$ is
connected to the $\rho_{00}$ dependence on $M_{p\omega}$, as different
intermediate baryon resonances with different masses dominate $\omega$
production in different $x_\mathrm{F}$ regions. The $\omega$ spin may hence be
differently aligned with different mother baryons.  

\begin{figure}[htp]
  \centering
\includegraphics[width=\textwidth]{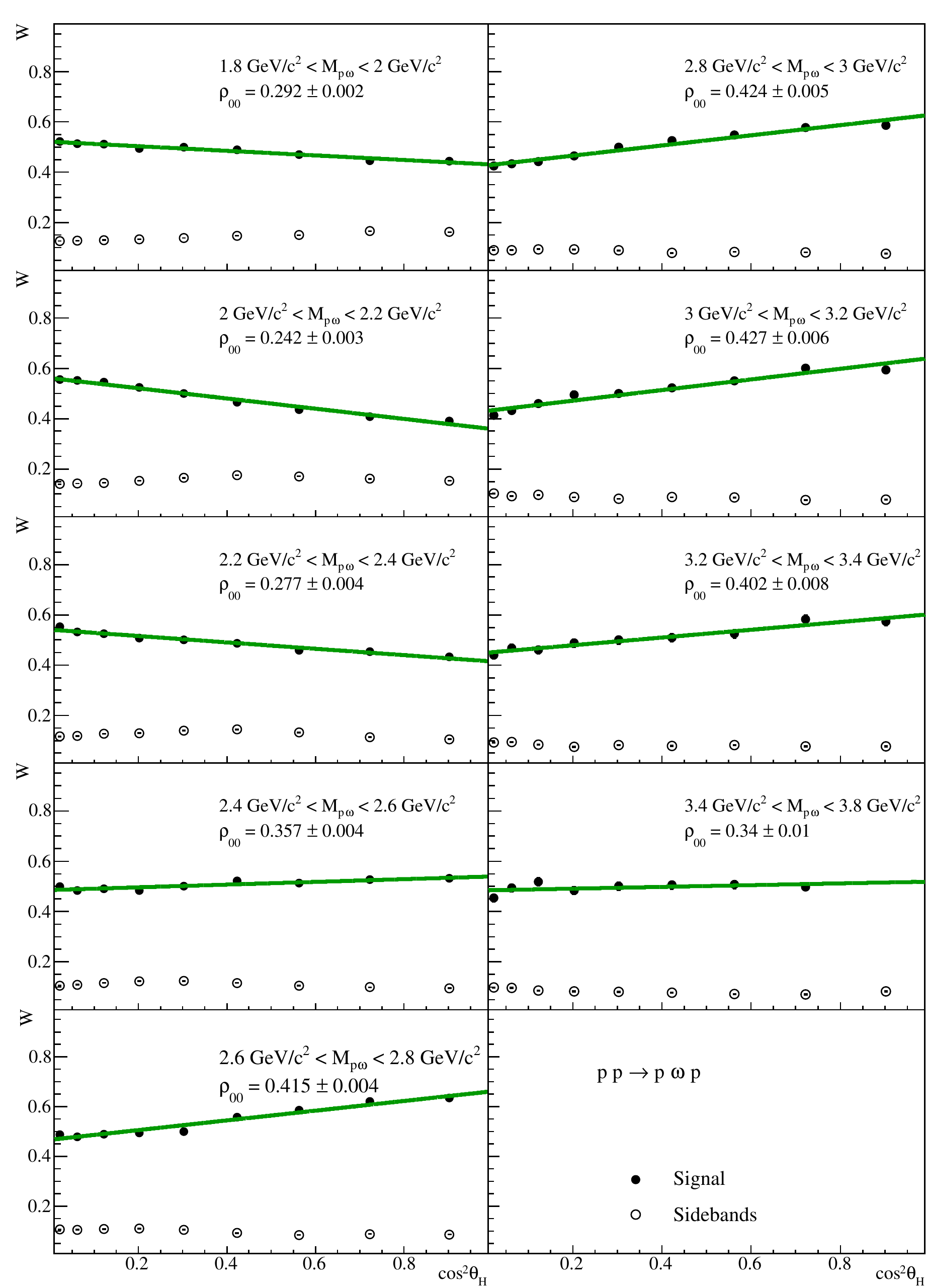}
  \caption[Helicity angle distributions]{Angular distributions of
    $\cos^{2}\theta$, where $\theta = \theta_H$ is the helicity angle of the
    $\omega$ meson in different $M_{p\omega}$ regions. From top left to bottom
    right, mass ranges in GeV/$c^2$: 1.8--2.0, 2.2--2.4, 2.4--2.6, 2.6--2.8,
    2.8--3.0, 3.0--3.2, 3.2--3.4, 3.4--3.8. The open points show the
    corresponding distribution for the events in the sidebands around the
    $\omega$ peak in the $M(\pi^+\pi^-\pi^0)$ distribution. The lines are the
    results of linear fits as explained in the text.} 
  \label{fig:cos2theta_Mpomega}
\end{figure}

\begin{figure}[htp]
  \centering
  \includegraphics[width=0.45\textwidth]{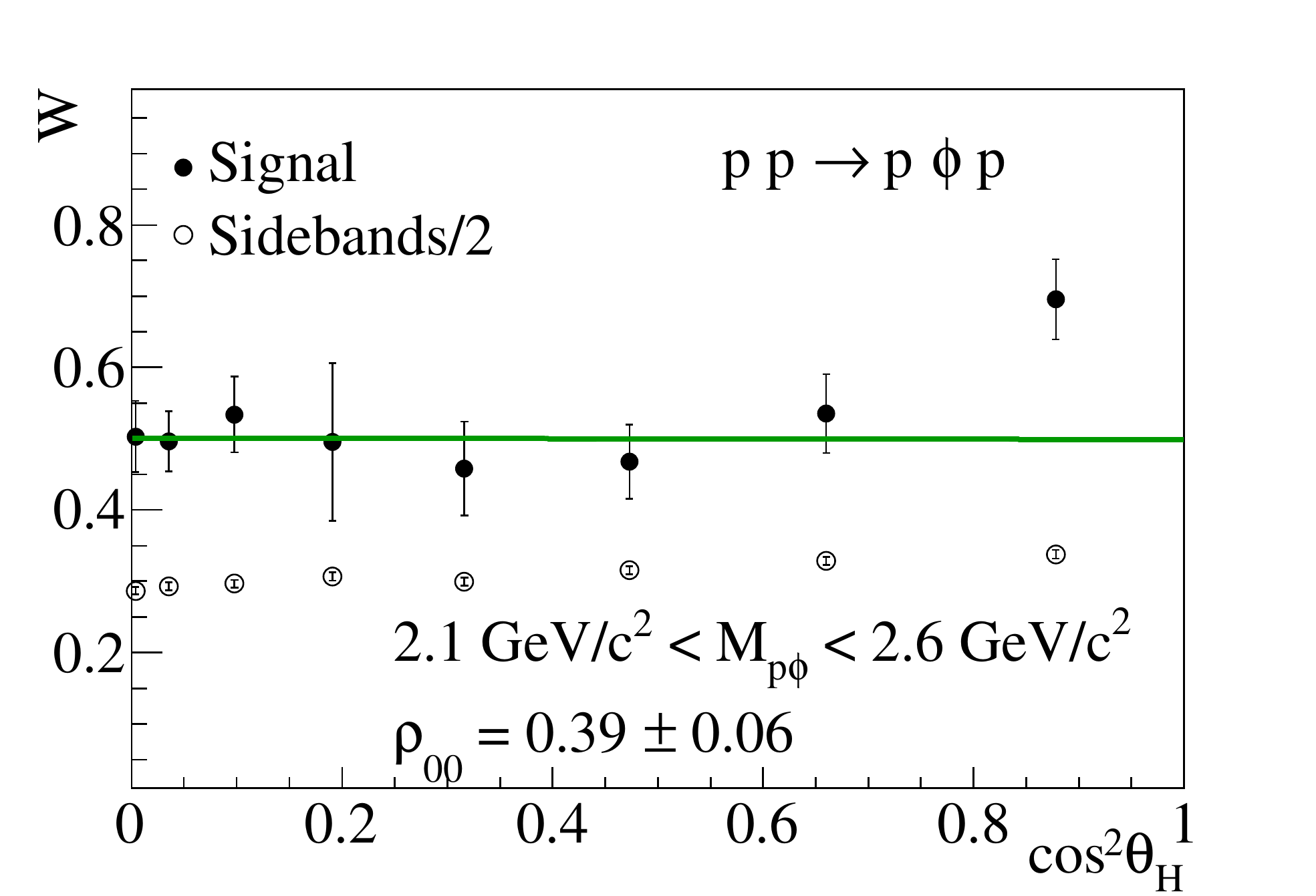}
  \includegraphics[width=0.45\textwidth]{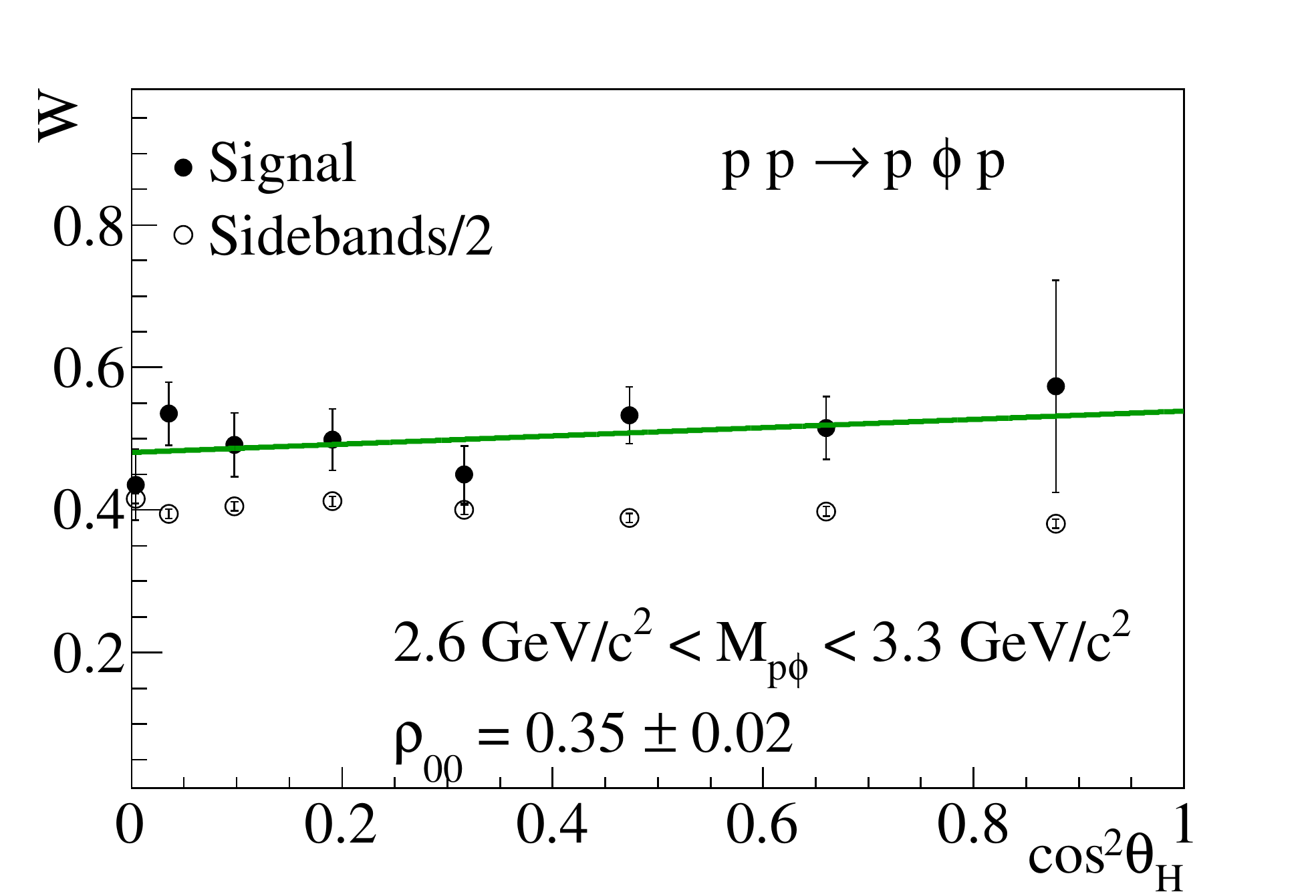}
  
  \caption[Helicity angle distributions]{Angular distributions of
    $\cos^{2}\theta$, where $\theta = \theta_H$ is the helicity angle of the
    $\phi$ meson for 2.1\,GeV/$c^2$ $ < M_{p\phi} < $ 2.6\,GeV/$c^2$ (left) and
    2.6\,GeV/$c^2$ $ < M_{p\phi} < $ 3.3\,GeV/$c^2$ (right). The open points
    show the corresponding distribution for the events in the sidebands around
    the $\phi$ peak in the $M(K^+K^-)$ distribution. The lines are the results
    of linear fits as explained in the text.} 
  \label{fig:cos2theta_Mpphi}
\end{figure}
\begin{table}[h]
	\caption{Upper section: Spin alignment $\rho_{00}$ extracted from the
          helicity angle distributions for $\omega$ production in the region
          $0.2 < x_\mathrm{F} < 0.9$ for different $M_{p\omega}$ regions. Middle
          section: The same but for $\phi$ production in the range $0.6 <
          x_\mathrm{F} < 0.9$. Lower section: The $\rho_{00}$ values extracted
          for $\omega$ within $0.6 < x_\mathrm{F} < 0.9$ and in the
          corresponding mass range as in the case of $\phi$ as explained in the
          text. The uncertainty is the propagated uncertainty from the linear
          fits, which in turn includes the quadratic sum of statistical
          uncertainties and uncertainties from the background subtraction.} 
\begin{center}
        \begin{tabular}[htp]{lcc}
          \hline
                Reaction & $M_{p\mathrm{V}}$ in GeV/$c^2$ & $\rho_{00}$ \\
                \hline
$pp \rightarrow pp \omega, \omega \rightarrow \pi^+\pi^-\pi^0$ & 1.8--2.0 & $0.292 \pm 0.002$ \\
$pp \rightarrow pp \omega, \omega \rightarrow \pi^+\pi^-\pi^0$ & 2.0--2.2 & $0.242 \pm 0.003$ \\
$pp \rightarrow pp \omega, \omega \rightarrow \pi^+\pi^-\pi^0$ & 2.2--2.4 & $0.277 \pm 0.004$ \\	
$pp \rightarrow pp \omega, \omega \rightarrow \pi^+\pi^-\pi^0$ & 2.4--2.6 & $0.357 \pm 0.004$ \\
$pp \rightarrow pp \omega, \omega \rightarrow \pi^+\pi^-\pi^0$ & 2.6--2.8 & $0.415 \pm 0.004$ \\
$pp \rightarrow pp \omega, \omega \rightarrow \pi^+\pi^-\pi^0$ & 2.8--3.0 & $0.424 \pm 0.005$ \\
$pp \rightarrow pp \omega, \omega \rightarrow \pi^+\pi^-\pi^0$ & 3.0--3.2 & $0.427 \pm 0.006$ \\
$pp \rightarrow pp \omega, \omega \rightarrow \pi^+\pi^-\pi^0$ & 3.2--3.4 & $0.402 \pm 0.008$ \\	
$pp \rightarrow pp \omega, \omega \rightarrow \pi^+\pi^-\pi^0$ & 3.4--3.8 & $0.35 \pm 0.01$ \\
\hline
$pp \rightarrow pp \phi, \phi \rightarrow K^+K^-$ & 2.1--2.6 & $0.39 \pm 0.06$ \\
$pp \rightarrow pp \phi, \phi \rightarrow K^+K^-$ & 2.6--3.3 & $0.35 \pm 0.02$ \\
\hline
$pp \rightarrow pp \omega, \omega \rightarrow \pi^+\pi^-\pi^0$ & 1.88--2.42 & $0.321 \pm 0.002$ \\	
$pp \rightarrow pp \omega, \omega \rightarrow \pi^+\pi^-\pi^0$ & 2.42--3.17 & $0.423 \pm 0.002$ \\
        \end{tabular}
\end{center}

	\label{tab:rho00M}
\end{table}

\begin{figure}[htp]
  \centering
  \includegraphics[width=0.9\textwidth]{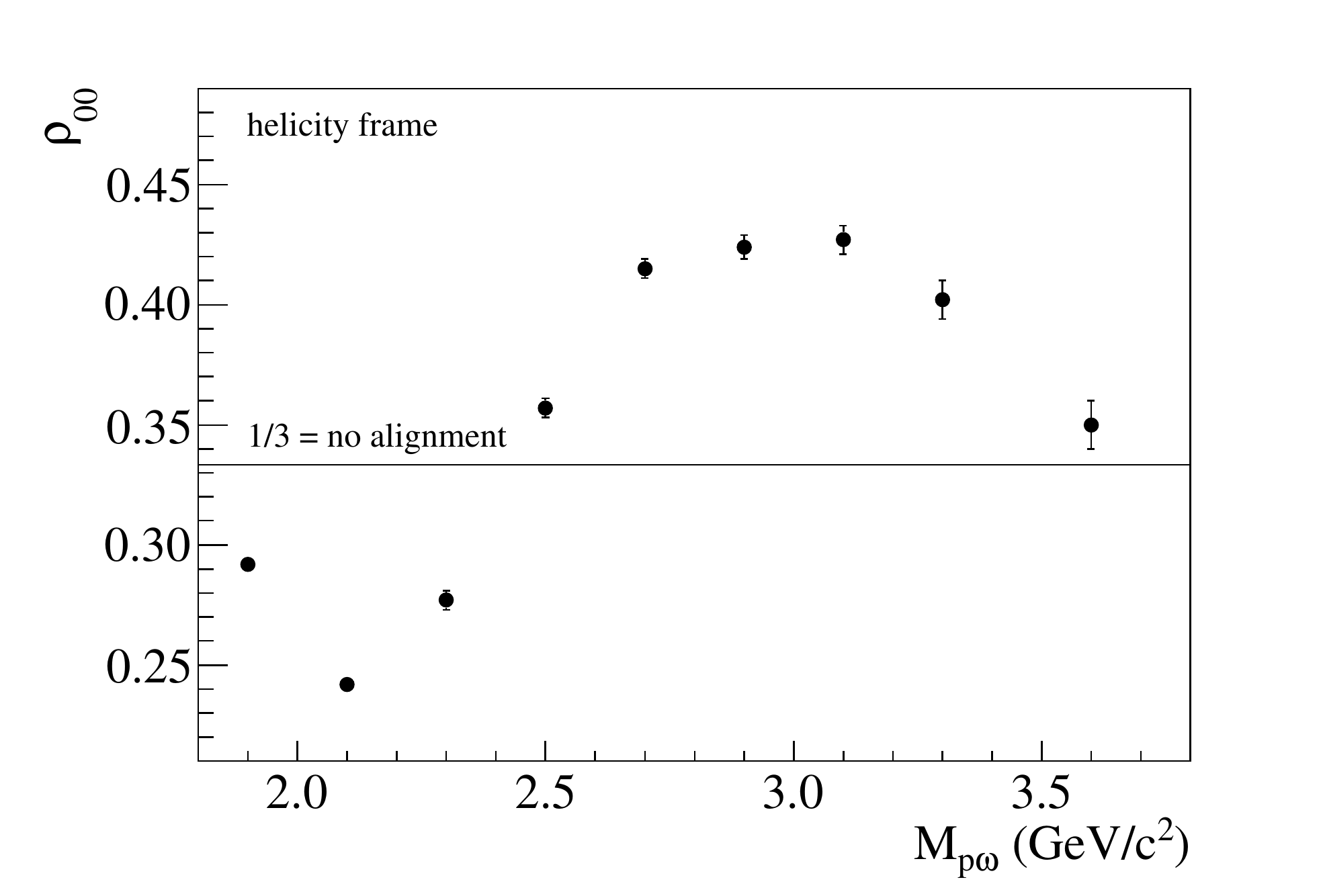}
 
  \caption[]{Spin alignment $\rho_{00}$ as a function of $M_{p\omega}$.}
  \label{fig:rho00M}
\end{figure}

The $M_{p\phi}$ spectrum (see Fig.\,\ref{fig:IM_pphi}) does not show apparent
structures and no baryon resonances are known to decay into
$p\phi$\,\cite{pdg}. This is in line with the $\rho_{00}$ results for $\phi$,
which are consistent with an unaligned $\phi$ with respect to a hypothetical
intermediate baryon, fairly independent of $x_\mathrm{F}$. The angular
distribution extracted in two different $M_{p\phi}$ ranges are both consistent
with isotropy. However, the errors are much larger than in the case of $\omega$
and a small alignment can therefore not be excluded.  In order to compare the
$\rho_{00}$ values from $\phi$ and $\omega$, we also extracted $\rho_{00}$ for
$\omega$ within the same $x_\mathrm{F}$ range and the corresponding
$M_{p\mathrm{V}}$ range as in the case of $\phi$. In the last four lines of
Table\,\ref{tab:rho00M}, the $M_{p\omega}$ and $M_{p\phi}$ ranges correspond to
the same $p_\mathrm{V}$ (see Eq.\,\ref{eq:p_V}) range. In the lower mass
intervals, the $\rho_{00}$ values agree within their combined errors, and the
difference is significant in the higher mass interval.  The high value of the
cross section ratio, the absence of structures in the $M_{p\phi}$ distribution,
the peaks in the $x_\mathrm{F}$ distributions in the lower-right panel of
Fig. \ref{fig:acc_xf} and the close-to-isotropic angular distributions indicate
that independent of $M_{p\phi}$ either a non-resonant diffractive process or a
central process dominates $\phi$ production within our kinematical range. Since
the COMPASS acceptance is small close to $M_{p\phi}$ = 2.1\,$\mathrm{GeV}/c^2$,
no conclusions can be drawn concerning the crypto-exotic $p\phi$ resonance
suggested in Ref.\,\cite{sibirtsevcrypto}.

\subsection{Spin alignment with respect to the transferred momentum}
\label{sec:spinex}
The isotropic $p\phi$ helicity angle distribution rises the question whether
there is a more natural choice of reference axis, to which also centrally
produced vector mesons are sensitive. Since both diffractive and central
production processes involve the exchange of at least one Reggeon, we define a
new reference axis by taking the direction of the momentum transfer from the
beam proton in the initial state to the fast proton in the final state, denoted
$\Delta\vec{P}$. In the rest system of the vector meson, this is opposite to the
momentum transfer from the target to the recoil. In the case of central
production, the dynamics of the vector meson should depend strongly on the
exchange, whereas in resonant diffractive production it is instead inherited
from the intermediate baryon resonance. The angle $\theta_{EX}$ is calculated in
the rest system of the vector meson with the same analyser as before.

The results are shown in Fig.\,\ref{fig:cos2theta_cp}. The extracted values of
$\rho_{00}$ are presented in Table\,\ref{tab:rho00cp} and in
Fig.\,\ref{fig:rho00cp}. The angular distribution of the background (open
circles / crosses) is isotropic, which demonstrates that the observed alignment
in the signal region is a real physical effect and not an artefact introduced by
the experiment.  Both $\phi$ and $\omega$ mesons are aligned transverse to the
direction of the exchanged Reggeon/Pomeron. The alignment is stronger when
$x_\mathrm{F}$ increases. In production processes without an intermediate state
or resonance, the vector meson will ``remember'' the direction of momentum
transfer of the incoming Pomeron, which in turn should influence the spin
orientation of the vector meson. This is the case in central production and when
the vector meson is produced by a shake-out of a $q\overline{q}$ object in the
proton.

The alignment of the $\omega$ meson reaches a maximum in the region
$0.7<x_\mathrm{F}<0.8$ while it is slightly smaller in the
$0.8<x_\mathrm{F}<0.9$.  The results for $\omega$ and $\phi$ show the same
trend, namely increasing anisotropy with increasing $x_\mathrm{F}$, and are
consistent with each other within uncertainties after removing the low-mass
resonant part of the $\omega$ data. This indicates that this reference axis is
only weakly sensitive to diffractive (resonant and non-resonant) production and
strongly sensitive to central production, as expected. Non-resonant diffractive
production (middle panel of Fig.\,\ref{fig:prodmech}) may contribute at low and
intermediate values of $x_\mathrm{F}$ while central production should dominate
at high $x_\mathrm{F}$.

\begin{figure}[htp]
  \centering
  \includegraphics[width=\textwidth]{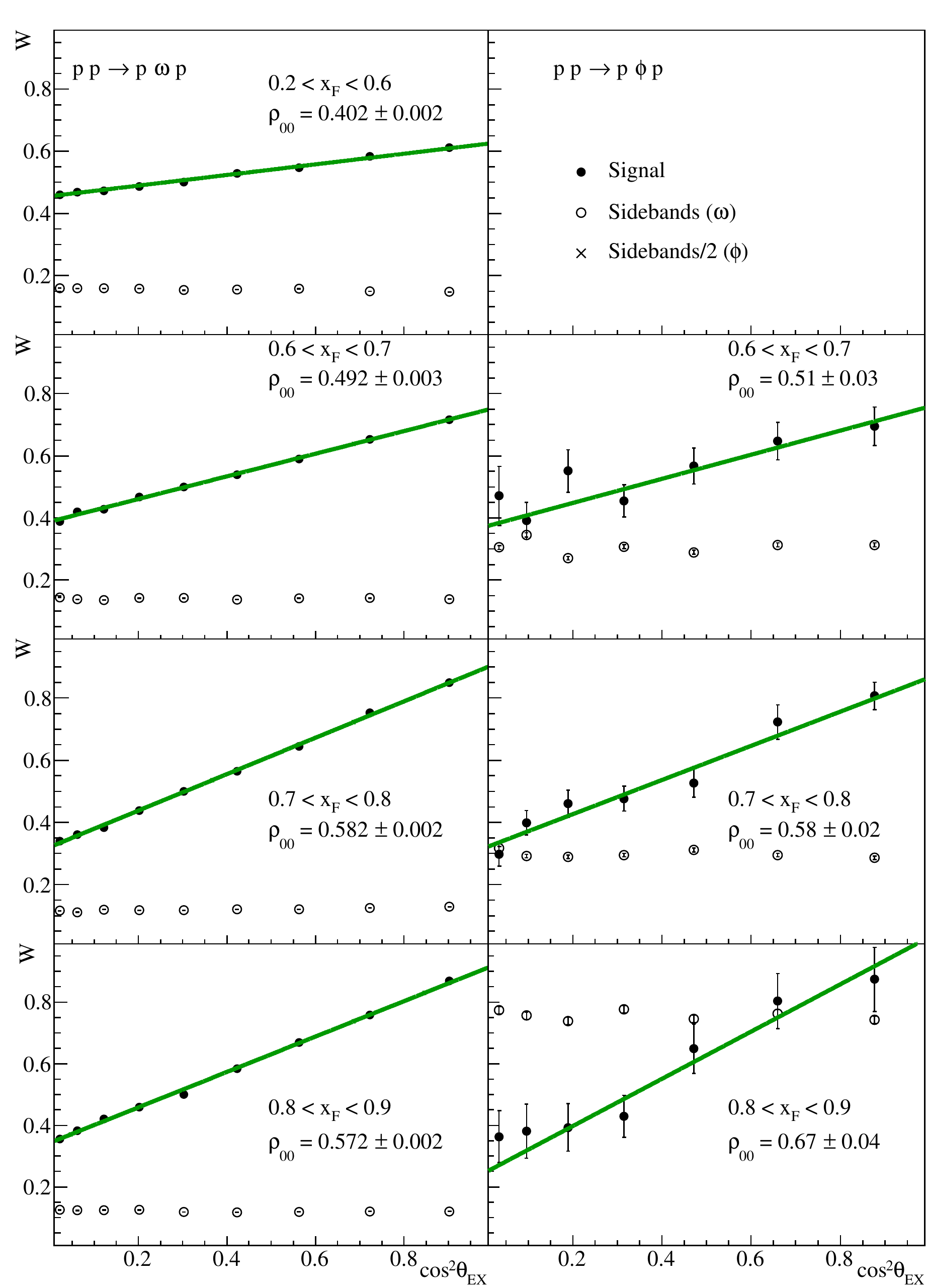}
  \caption[Angle w.r.t. $\Delta\vec{P}$]{Angular distributions with respect to
    $\cos^{2}\theta = \cos^{2}\theta_{EX}$ using the momentum transfer from
    $p_\mathrm{beam}$ to $p$, $\Delta\vec{P}$, as reference axis. The panels
    show different $x_\mathrm{F}$ regions: 0.2--0.6 (top), 0.6--0.7 (second
    line), 0.7--0.8 (third line) and 0.8--0.9 (bottom). The error bars include
    statistical errors and systematics from the background subtraction. The open
    points show the corresponding distribution for the events in the sidebands
    around the $\omega$ peak in the $M(\pi^+\pi^-\pi^0)$ distribution. The
    crosses show the corresponding distribution (scaled by 0.5) for the events
    in the sidebands around the $\phi$ peak in the $M(K^+K^-)$ distribution. The
    lines are the results of linear fits as explained in the text.}
  \label{fig:cos2theta_cp}
\end{figure}

\begin{table}[h]
	\caption{Spin alignment $\rho_{00}$ extracted using $\Delta\vec{P}$ as
          reference axis. The Table includes $\phi$ and $\omega$ production. The
          results for different $p_\mathrm{V}$ cuts are also given for
          $\omega$. The uncertainty is the propagated uncertainty from the
          linear fits, which in turn includes the quadratic sum of statistical
          uncertainties and uncertainties from the background subtraction.} 
\begin{center}
        \begin{tabular}[htp]{lcc}
          \hline
                Reaction & $x_\mathrm{F}$ & $\rho_{00}$ \\
                \hline

$pp \rightarrow pp \phi, \phi \rightarrow K^+ K^-$ & 0.6--0.7 & $0.51 \pm 0.03$ \\
$pp \rightarrow pp \phi, \phi \rightarrow K^+ K^-$ & 0.7--0.8 & $0.58 \pm 0.02$ \\
$pp \rightarrow pp \phi, \phi \rightarrow K^+ K^-$ & 0.8--0.9 & $0.67 \pm 0.04$ \\
\hline
$pp \rightarrow pp \omega, \omega \rightarrow \pi^+\pi^-\pi^0$ & 0.2--0.6 & $0.408 \pm 0.002$ \\
$pp \rightarrow pp \omega, \omega \rightarrow \pi^+\pi^-\pi^0$ & 0.6--0.7 & $0.492 \pm 0.003$ \\	
$pp \rightarrow pp \omega, \omega \rightarrow \pi^+\pi^-\pi^0$ & 0.7--0.8 & $0.582 \pm 0.002$ \\
$pp \rightarrow pp \omega, \omega \rightarrow \pi^+\pi^-\pi^0$ & 0.8--0.9 & $0.572 \pm 0.002$ \\
\hline
$pp \rightarrow pp \omega, \omega \rightarrow \pi^+\pi^-\pi^0$, $p_\mathrm{V} > 1.0$\,GeV/$c$ & 0.6--0.7 & $0.39 \pm 0.01$ \\	
$pp \rightarrow pp \omega, \omega \rightarrow \pi^+\pi^-\pi^0$, $p_\mathrm{V} > 1.0$\,GeV/$c$ & 0.7--0.8 & $0.527 \pm 0.005$ \\
$pp \rightarrow pp \omega, \omega \rightarrow \pi^+\pi^-\pi^0$, $p_\mathrm{V} > 1.0$\,GeV/$c$ & 0.8--0.9 & $0.577 \pm 0.002$ \\
\hline
$pp \rightarrow pp \omega, \omega \rightarrow \pi^+\pi^-\pi^0$, $p_\mathrm{V} > 1.4$\,GeV/$c$ & 0.8--0.9 & $0.601 \pm 0.005$ \\
        \end{tabular}
\end{center}

	\label{tab:rho00cp}
\end{table}

\begin{figure}[htp]
  \centering
  \includegraphics[width=0.9\textwidth]{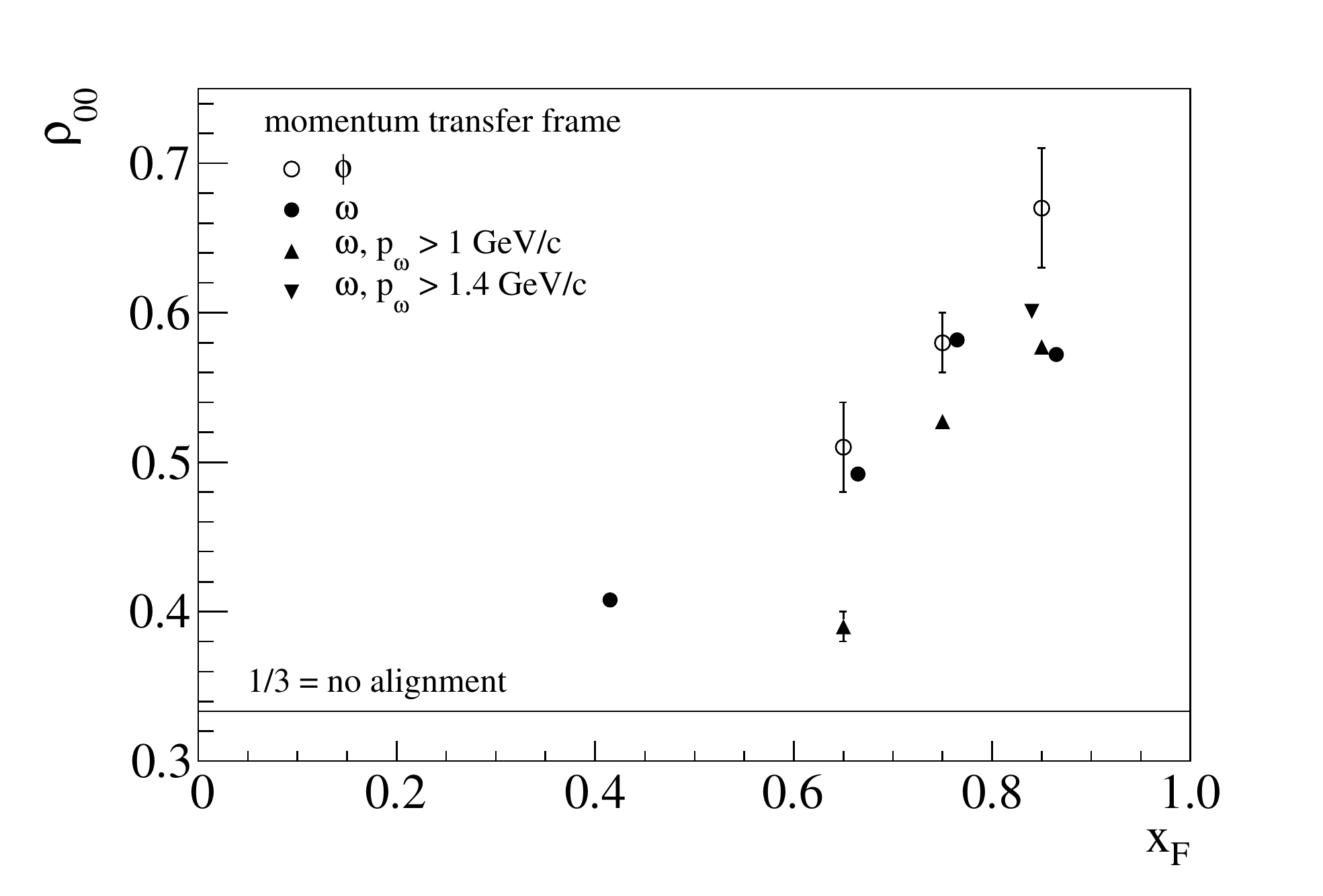}
  \caption{Spin alignment $\rho_{00}$ extracted using $\Delta\vec{P}$ as reference axis as a function of $x_\mathrm{F}$ for different $p_\mathrm{V}$ cuts.}
 \label{fig:rho00cp}
\end{figure}

\section{Discussion}
\label{sec:nonres}
An important process in exclusive $\omega$ meson production appears to be
diffractive excitation of the beam proton with the excitation into nucleon
resonances followed by a two-body decay $N^{*}\rightarrow p\omega$. This is
supported by the structures in the $M_{p\omega}$ spectra in
Figs.\,\ref{fig:IM_pphi} and \,\ref{fig:IM_pomega3}, which are consistent with
known high-spin resonances\,\cite{pdg}, and the significant alignment of the
$\omega$ meson with respect to the direction of the $p\omega$ system. The
alignment is strongly dependent on $M_{p\omega}$. The $N^{*}$ spin is aligned
with its direction. In a two body decay, high spin resonances have to emit the
vector meson with an orbital angular momentum, $\vec{J} = \vec{L} + \vec{J}_p +
\vec{J}_\mathrm{V}$.  If the vector meson spin is preferentially aligned with
the direction of the orbital angular momentum, then we expect an increasing
anisotropy of the vector meson decay in the helicity frame of the $N^{*}$ with
increasing spin of the resonance.

The fact that no structures are visible in the $p\phi$ spectrum and the
observation that the $\phi$ meson is unaligned in the $p\phi$ helicity system
indicates that $N^{*}$ decays into $p\phi$ are OZI suppressed, reflecting the
internal structure of the resonance. The observed violation of the OZI rule by a
factor of 3-4 (see Table\,1) indicates either an admixture of other,
OZI-violating reaction processes or a genuine violation of the predicted
$g_{\phi NN}^{2}/g_{\omega NN}^{2}$ coupling ratio. Note that similar and
sometimes smaller values of the OZI violation factor (about 2-3) were observed
in Refs.\,\cite{sphinx1,baldini,baldi,blobel}, all in a kinematic domain where
$N^{*}$ production is prominent.

Removing the low-mass region with visible resonances by a cut on the vector
meson momentum in the $pV$ rest system, $p_\mathrm{V} > 1.4$\,GeV/$c$,
\textit{i.e.} $M_{p\omega} > 3.3$\,GeV/$c^2$, the picture changes
significantly. The $\omega$ spin is found to be unaligned with respect to the
$p\omega$ system, consistent with the absence of resonances. Furthermore, the
OZI violation increases and converges to a factor of about 8, independently of
$x_\mathrm{F}$, as can be seen in Table\,2. This is in remarkable agreement not
only with the SPHINX analysis\,\cite{sphinx1} after removal of the
low-$M_{p\omega}$ region, but surprisingly also with data close to threshold
from ANKE\,\cite{hartmann}, DISTO\,\cite{balestra} and
COSY-TOF\,\cite{elsamad,abdelbary2}.

The high mass part of the $M_{p\mathrm{V}}$ spectrum shows no structures, but
may still contain $N^{*}$ resonances which probably are broad and largely
overlap. The angular distributions are isotropic, which means that either
low-spin resonances contribute, which is however unlikely in this mass region,
or the contribution of resonances is small.

In the high-mass continuum, the decays of $\omega$ and $\phi$ mesons are both
strongly aligned with the direction of the 3-momentum transfer
$\Delta\vec{P}$. The similar behaviour of the alignments together with larger
$\rho_{00}$ values with increasing $x_\mathrm{F}$ indicates that the production
mechanism is the same for $\omega$ and $\phi$ in this region. This may point to
a central Pomeron--Reggeon fusion which produces a vector meson. The OZI
violation then reflects a hidden flavour-flow with the emitted Reggeon. The
observed $x_\mathrm{F}$ dependence of $\rho_{00}$ with respect to
$\Delta\vec{P}$, where $\rho_{00}$ increases with increasing $x_\mathrm{F}$,
suggests this process since central production favours large $x_\mathrm{F}$ of
the fast proton. A different approach to this reaction is obtained assuming an
alignment of the spins of the vector meson with the angular momentum of its
emission with respect to $\Delta\vec{P}$. Then, the transferred angular momentum
has to be perpendicular to $\Delta\vec{P}$. We can regard these events as
scattering off a Pomeron radiated from the target proton and absorbed by a
colourless object in the beam proton wave function, which carries some fraction
of the total momentum. This kind of mechanism may be associated with
non-resonant diffractive dissociation. In a very simple picture, the proton
dissociates into a proton plus a virtual (off-shell) vector meson $V^*$ (in
Ref.\,\cite{ellis}, this process is referred to as a shake-out). If the Pomeron
emitted from the target recoil proton is absorbed by $V^*$, this could result in
an on-shell vector meson recoiling along the direction of momentum transfer of
the Pomeron.  In other words, we expect that at some energy scale the Pomeron
should resolve structures in the extended proton.  The data show evidence for
this in the observed angular distributions of the vector meson decays, as shown
in Figs.\,10 and 11 and summarised in Table\,5. They exhibit large anisotropies
increasing with $x_{F}$, which indicates the presence of a transversely
localised process with a dependence of its direction on $\Delta\vec{P}$. The
high OZI violation indicates a higher effective resolution scale in this process
and reflects the probability of finding a preformed $\phi$ meson relative to the
preformed $\omega$ meson at a resolution scale near $m_{\phi}\approx
1$\,GeV/$c^2$. The natural angular momentum quantisation axis for such a process
is the direction of the momentum transfer mediated by the Pomeron. Both $\omega$
and $\phi$ have substantial alignment of their spins perpendicular to this axis,
indicating a transferred orbital angular momentum. The latter is naturally
oriented perpendicular to the direction of momentum transfer to which the
angular momentum of the vector mesons has a tendency to align if spin-orbit
forces occur.

It has been already noted that Pomeron-Pomeron fusion into a $J^{PC}(I^{G})$ =
$1^{--}(0^{-})$ meson is forbidden due to $G$-parity conservation. Another
theoretical possibility is a central Pomeron-Odderon process
\footnote{An Odderon is similar to the Pomeron but with negative parity, charge
  conjugation and $G$-parity.}. Since this process involves no quark lines and
the only difference between $\omega$ and $\phi$ is the mass, the $\phi$
production rate should be of the same order as the $\omega$ rate. This is in
sharp contrast to our data, in which the $\omega$ cross section is thirty times
larger than that of the $\phi$. Our data therefore show no evidence for
Pomeron-Odderon fusion in our kinematic domain ($\sqrt{s}$=18.97\,GeV,
$0.1<t'<1.0$\,(GeV$/c)^2$).

\section{Summary and Conclusion}
In this work, exclusive $\phi$ and $\omega$ vector meson production in the
reaction $pp\rightarrow pVp$ has been measured. We find OZI violations ranging
from $F_\mathrm{OZI}=3$ to $F_\mathrm{OZI}=9$ depending on the kinematic region.
The invariant mass $M_{p\mathrm{V}}$ of the forward proton and the vector meson
appears to be the most important kinematic quantity in our study to discriminate
processes with different mechanisms.  The clear structures in the $M_{p\omega}$
spectrum indicate the importance of $pp \rightarrow pN^{*}, N^{*} \rightarrow
p\omega$ in $\omega$ production. This is also supported by the significant
alignment of the spin of the $\omega$ meson with respect to the direction of the
$p\omega$ system. In the case of decays into a ground state vector meson, the
$N^{*}$ has to transfer considerable angular momentum.  The absence of
structures in the $M_{p\phi}$ spectrum in combination with no observed alignment
of the $\phi$ spin with respect to the direction of the $p\phi$ system shows
that the decay of the $N^{*}$ resonances into $p\phi$ is OZI suppressed. This
indicates that the $s\overline{s}$ component of such resonances must be very
small. The observed OZI violation by a factor 3-4 in this region could be either
due to the admixture of other processes or a genuine violation of the predicted
$g^2_{\phi NN}/g^2_{\omega NN}$ ratio.

Removing the resonance region by requiring $M_{p\omega} > 3.3$\,GeV/$c^2$, the
OZI violation in the remaining kinematic range is significantly higher,
typically of order $8\pm1$. Moreover, the spin of both $\omega$ and $\phi$ are
unaligned with respect to the $pV$ system. The behaviour of both vector mesons
is the same in the system defined by the transferred momentum. This indicates
that the production mechanism in this region for both $\omega$ and $\phi$ is
central Reggeon--Pomeron fusion, with the observed OZI violation reflecting a
hidden flavour flow. This process can also be regarded as a Pomeron resolving
preformed colourless objects in the proton wave function and ejecting them in a
shake-out. The direction of the transferred momentum is remembered by the vector
meson and is manifested in its decay angular distributions. The OZI violation
then reflects the probability of resolving a $s\overline{s}$ state in the
nucleon.

\section*{Acknowledgements}
We are grateful to Prof. Colin Wilkin and Prof. Stefan Leupold for helpful
discussions. We also gratefully acknowledge the support of the CERN management
and staff and the skill and effort of the technicians of our collaborating
institutes. Special thanks go to V. Pesaro for his technical
support during the installation and the running of this experiment. This work
was made possible by the financial support of our funding agencies.

\end{document}